\documentclass[%
 reprint,
 amsmath,amssymb,
 aps,
]{revtex4-1}
\usepackage{float}
\usepackage{graphicx}
\usepackage{dcolumn}
\usepackage{bm}
\usepackage{hyperref}
\newtheorem{theorem}{Theorem}

\newtheorem{proposition}{Proposition}

\usepackage{color}

\begin{document}

\preprint{APS/123-QED}

\title{Fault-tolerant quantum speedup from constant depth quantum circuits}

 \author{Rawad Mezher $^1$$^{,2,3}$}
 \email{rmezher@exseed.ed.ac.uk}
 
 \author{Joe  Ghalbouni $^2$} \author{Joseph Dgheim $^2$} \author{Damian Markham $^{1,4}$}
\email{damian.markham@lip6.fr}
 \affiliation{(1) Laboratoire d'Informatique de Paris 6, CNRS, Sorbonne Universit\'e, 4 place Jussieu, 75252 Paris Cedex 05, France}
 \affiliation{(2) Laboratoire de Physique Appliquée, Faculty of Sciences 2, Lebanese University, 90656 Fanar, Lebanon}
 \affiliation{(3) School of Informatics, University of Edinburgh, 10 Crichton Street, Edinburgh, EH8 9AB}
 \affiliation {(4) JFLI, National Institute for Informatics, and the University of Tokyo, 
 Tokyo, Japan.}

\date{\today}

\begin{abstract}

A defining feature in the field of quantum computing is the potential of a quantum device to outperform its classical counterpart for a specific computational task. 
By now, several proposals exist showing that certain sampling problems can be done efficiently quantumly, but are not possible efficiently classically, assuming strongly held conjectures in complexity theory. A feature  dubbed \emph{quantum speedup}. However, the effect of noise on these proposals is not well understood in general, and in certain cases it is known that simple noise can destroy the quantum speedup.

Here we develop a fault-tolerant version of one family of these sampling problems, which we show can be implemented using  quantum circuits of constant depth.
We present two constructions, each taking $poly(n)$ physical qubits, some of which are prepared in noisy magic states. 
The first of our constructions is a constant depth quantum circuit composed of single and two-qubit nearest neighbour Clifford gates in four dimensions. This circuit has one layer of interaction with a classical computer before final measurements.
Our second construction is a constant depth quantum circuit with single and two-qubit nearest neighbour Clifford gates in three dimensions, but with two layers of interaction with a classical computer before the final measurements.

For each of these constructions, we show that there is no classical algorithm which can sample according to its output distribution in $poly(n)$ time, assuming two standard complexity theoretic conjectures hold. The noise model we assume is the so-called \emph{local stochastic quantum noise}. 
Along the way, we introduce various new concepts such as constant depth magic state distillation (MSD), and constant depth output routing, which arise naturally in measurement based quantum computation (MBQC), but have no constant-depth analogue in the circuit model.


\bigskip

 
 

\end{abstract}
\maketitle
\bigskip
\textbf{\emph{Introduction }}- Quantum computers promise incredible benefits over their classical counterparts in various areas, from breaking RSA encryption \cite{schor94}, to machine learning \cite{biamonte2017quantum}, and improvements to generic search \cite{grover96}, among others \cite{montanaro2016quantum,olson2017quantum}.
 Although these and other examples of quantum algorithms do outperform classical ones, on the practical level, they in general require quantum computers with a high level of fault-tolerance and scalability,  the likes of which appear to be out of the reach of current technological developments \cite{Preskill18}.  
 An interesting question is thus, what can be done with so-called \emph{sub-universal} quantum devices which are not universal, in the sense that they cannot perform any quantum computation, but are realizable in principle by our current technologies.
 Several examples of such practically motivated sub-universal models which nevertheless capture a sense of quantum advantage have been discovered in recent years \cite{BJS10,AA11,BMS16PRL,HM18,gao17,BHS+17,HBS+17,MB17,MGDM19,BIS+18,NRK+18,AAB+19,NBG+19,HHB+19,BGK+19,bravyi2018quantum}. 
 In most of these works, sampling from the output probability distribution of these sub-universal devices has been shown to be classically impossible to do efficiently, provided widely believed complexity theoretic conjectures hold \cite{BJS10,AA11}. 
 Thus, these devices demonstrate what is known as an exponential $quantum$ $speedup$. 
 
 The first experimental demonstration of quantum speedup is a major milestone in quantum information. Recent audacious experimental efforts \cite{AAB+19} and subsequent proposals of their classical simulation \cite{IBM} bring to light the challenges and subtleties of achieving this goal. Statements of quantum speedup are complexity theoretic in nature, making it difficult to pin down when a problem can in practice be simulated or not  classically, even if we know in the limit of `infinite size' experiments that efficient classical simulation is impossible. At the same time, the role of noise in simplifying the simulation is ever more important, as systems grow, noise becomes more difficult to control, and it is a subtle question as to when it dominates; and even simple noise can very easily lead to breakdown of quantum speedup. Indeed, in \cite{BMS16,OB18,shchesnovich2019noise,TTT+20,KLF20,yung2017can,gao2018efficient} it was shown that noise generally renders the output probabilities of these devices (which in the noiseless case demonstrate quantum speedup) classically simulable efficiently.
There is clearly a great need to understand better the effect of noise, and develop methods of mitigation.
 
 Applying the standard approach to deal with noise in computation, fault-tolerance, is non-trivial in this setting for at least two reasons \cite{fowler2012proof,NC2000,LAR+11,DKL+02}. Firstly, the resources it consumes can be huge. 
 Secondly, it typically involves operations that step outside of the simplified computational model that makes it attractive in the first place.
For example, in \cite{BJS10} the sub-universal model IQP was defined, as essentially the family of circuits where all gates are diagonal in the $X$-basis, and shown to provide sampling problems demonstrating quantum speedup in the noiseless case. However in \cite{BMS16} it was shown that a simple noise model - each output bit undergoes a bit flip with probability $\varepsilon$ - renders the output probabilities of sufficiently anti-concentrated IQP circuits efficiently simulable classically. Interestingly, for this special type of noise, they also show that quanutm speedup can be recovered using classical fault-tolerance and larger encodings of the problem quantumly, still within the IQP framework \cite{BMS16}. However, for more general noise (for example Pauli noise in all the Pauli bases), this does not appear to work, and it is not obvious if it is possible to do so within the constrained computational mode. In this case that would mean maintaining all gates be diagonal in $X$, which is not obvious as typical encoding and syndrome measurements involve more diverse gates.


In this work, we study how quantum speedup can be demonstrated in the presence of noise for a family of sampling problems. 
We take the \emph{local stochastic quantum noise} (we will also refer to this noise as local stochastic noise) model, commonly studied in the quantum error correction and fault-tolerance literature \cite{BGK+19,FGL18,gottesman2013fault,aliferis2005quantum,aliferis2007accuracy}.
Our sampling problems are built on a family of schemes essentially based on local measurements on regular graph states, which correspond to constant depth 2D nearest neighbor quantum circuits showing quantum speedup \cite{gao17,BHS+17,HBS+17,MGDM18,MGDM19,HHB+19}.
We show that these can be made fault-tolerant in a way which maintains constant depth of the quantum circuits, albeit with large (but polynomial) overhead in the number of ancilla systems used, and at most two rounds of (efficient) classical computation during the running of the circuit.

We present two different constructions based on two different techniques of fault-tolerance, the first of which involves the use of transversal gates and topological codes each encoding a single logical qubit \cite{BGK+19,fowler2012proof,wang2011surface}. This construction results in a constant depth quantum circuit demonstrating a quantum speedup, but, because of the need for long range transversal gates, can only be viewed as a quantum circuit with single qubit Clifford gates and  nearest  neighbor two-qubit Clifford gates in 4D (we will henceforth refer to this as our 4D nearest neighbor (NN)  architecture). Our  second construction avoids using transversal gates by exploiting topological defect-based quantum computation \cite{RHG07}, thereby resulting in a constant depth quantum circuit which is a  3D NN architecture. The tradeoff, unfortunately, is that our 3D NN architecture requires polynomially more ancillas than our 4D NN architecture, and has two layers of interaction with a classical computer, as compared to one such layer in our 4D NN architecture.

Our first construction in 4D uses several techniques from \cite{BGK+19}, in particular regarding the propagation of noise through Clifford circuits. For the second construction, we also develop techniques from \cite{KT19}.
In \cite{KT19}, a construction for fault-tolerant quantum speedup was presented which consisted of a constant depth quantum circuit obtained by using defect-based topological quantum computing \cite{RHG07}. This construction is non-adaptive (no interaction with classical computer during running of circuit), and can be viewed as a 3D NN architecture. The main disadvantage of the construction in \cite{KT19} was the magic state  distillation (MSD) procedure employed, which makes the scheme impractical in the sense that one should repeat the experiment an exponential number of times in order to observe an instance which is hard for the classical computer to simulate. In both our 3D and 4D NN constructions, we overcome this problem by optimizing our MSD procedure, thereby making the appearance of a hard instance very likely in only a few repetitions of the experiment, a feature called single-instance hardness \cite{gao17}. This, however, comes at the cost of adding adaptive interactions with the classical computer while running the quantum circuit. 

This paper is  organised as follows. First, we introduce the family of sampling problems using graph states, on which our constructions are based. After briefly defining the noise model, we describe in detail the encoding procedure for our 4D NN architecture. We then describe the effects of noise on our construction, step by step, starting from the Clifford part of the circuit and ending with the MSD, while introducing our optimized MSD techniques based on MBQC, namely constant depth non-adaptive MSD, and MBQC routing.   Finally, we explain how to modify, using our optimized MSD techniques, the 3D NN architecture in \cite{KT19} in order to give rise to the single-instance hardness feature \cite{gao17}. Note that in our 3D NN architecture, we use different (fixed) measurement angles to those in \cite{KT19} to construct a different sampling problem having an anti-concentration property \cite{MGDM18,MB17,HBS+17}.


\bigskip

\textbf{\emph{Graph state sampling}} - Our approach is to construct a fault-tolerant version of the architectures based on measurement based quantum computation (MBQC) \cite{RB01}, which have recently been shown to demonstrate a quantum speedup \cite{gao17,BHS+17,HBS+17,MGDM18,MGDM19,HHB+19}. 
In these constructions, the sampling is generated by performing local measurements on a large entangled state, known as a graph state.
Given a graph $G$, with vertices $V$ and edges $E$, the associated graph state $|G\rangle$, of $|V|$ qubits is defined as
\begin{equation}
    \label{eq1PRL}
    |G\rangle:=\prod_{\{i,j\} \in E}CZ_{ij}\bigotimes_ {a \in V}|+\rangle_a,
\end{equation}
where $|+\rangle:=\dfrac{|0\rangle+|1\rangle}{\sqrt{2}}$ and $CZ_{ij}$ is the controlled-Z gate ($CZ$) acting on qubits $i$ and $j$ connected by an edge. 
For certain graphs of regular structure, such as the cluster \cite{RB01} or brickwork \cite{BFK+09} states, applying single qubit measurements, of particular choices of angles on the $XY$-plane, effectively samples distributions, in a way that is impossible to do efficiently classically, up to the standard assumptions \cite{BHS+17,HBS+17,gao17,MGDM18,KT19}.

Although our techniques can be applied to $any$ such architecture where the measurement angles in the $XY$-plane of the Bloch sphere are chosen from  the set $\left\{0,\dfrac{\pi}{2},\dfrac{\pi}{4}\right\}$ \cite{gao17,BHS+17,HBS+17,MGDM18}; for concreteness we will focus on the architecture of \cite{MGDM18}. 

\begin{figure*}[]
\begin{center}
\graphicspath{}
\includegraphics[scale=0.6]{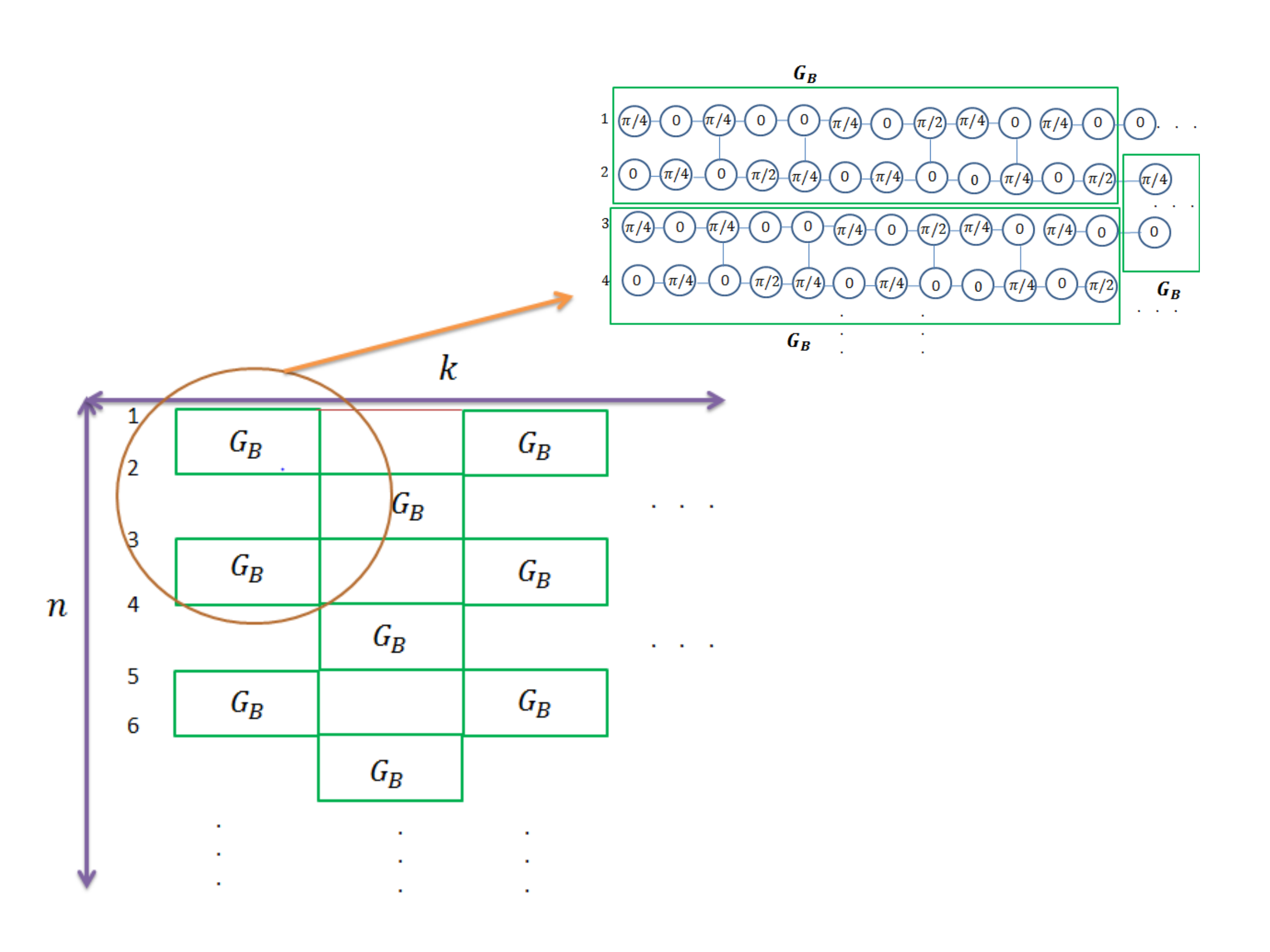}
\caption{   Graph state $|G\rangle$ of \cite{MGDM18} together with the pre-specified measurements in the $XY$ plane.  This graph state is composed of $n$ rows and $k$ columns as seen in the main text (lower part of figure), and made up of two-qubit gadgets $G_B$ (green rectangles) zoomed in at the upper part of the figure (orange circle and arrow). Blue circles are qubits, blue vertical and horizontal lines are $CZ$ gates, the symbols inside each circle correspond to the angle in the $XY$ plane at which this qubit is measured. The  $\pi/4$ symbol  is a measurement at an angle $\pi/4$ in the $XY$ plane, similarly for  $\pi/2$ and $0$. 
In the original construction of \cite{MGDM18}, the red horizontal line is a long range $CZ$, these are used periodically in $|G\rangle$ to connect two consecutive $G_B$ gadgets acting on qubits of either the first row or the last row of $|G\rangle.$ Here, this red horizontal line is a linear cluster of twelve qubits measured at an $XY$ angle of 0, this is in order to make the construction nearest neighbor. Note that this only adds single qubit random Pauli gates to the random gates of \cite{MGDM18}, and therefore does not affect their universality capacity in implementing a $t-$design.}
\label{FIG G original}
\end{center}
\end{figure*}

Following \cite{MGDM18} we start with a regular graph state, closely related to the brickwork state \cite{BFK+09}, composed of $n$ rows and $k$ columns.
Then we (non-adaptively) measure qubits of all but the last column at pre-specified fixed $XY$-angles from the set $\left\{0,\dfrac{\pi}{2},\dfrac{\pi}{4}\right\}$ effectively applying a unitary, on the $n$ unmeasured qubits. This is illustrated in Figure \ref{FIG G original}.

Let  $V_1 \subset V$  be the set of qubits which are measured at angle $\dfrac{\pi}{4}$ and $V_2 \subset V$ is the set of qubits which are measured at an $XY$ angle $\dfrac{\pi}{2}$. One can equivalently perform local rotations to the graph state and measure all systems in the $Z$ basis. In this way, if we define 
\begin{eqnarray}
|G'\rangle:= \left(\bigotimes_{a \in V}H_a \bigotimes_{b\in V_1} Z_b\left(\pi/4\right)\bigotimes_{c\in V_2}Z_c\left(\pi/2\right) \right)|G\rangle, \nonumber 
\end{eqnarray}
where $H$ is the Hadamard unitary and $Z(\theta):=e^{-i{\dfrac{\theta}{2} Z}}$ is a rotation by $\theta$ around Pauli Z, then one can represent the outcome by a measurement result bit string $s\in \{0,1\}^{n.(k-1)}$, with associated resultant state
\begin{eqnarray}
    \label{eq2PRL}
 \langle s|G'\rangle    =  \dfrac{1}{\sqrt{2^{n.(k-1)}}} U_s|0\rangle^{\otimes n}.
\end{eqnarray}

This procedure effectively samples from the ensemble of unitaries $\left\{\dfrac{1}{2^{n.(k-1)}},U_s\right\}.$ It was shown in \cite{MGDM18} that setting $k=O(t^{9}(nt+log(\dfrac{1}{\varepsilon})))$, this ensemble has the property of being an $\varepsilon$-approximate unitary $t$-design \cite{DCE+09} - that is, it approximates sampling on the Haar measure up to the $t$-th moments. This property allows us to reduce the requirements for the proof of quantum speedup since it implies anti-concentration for $t=2$ from \cite{HBS+17}.

Measuring qubits of the last column in the computational ($Z$) basis and denoting the outcome by a bit string $x \in \{0,1\}^n$, our construction samples the bit strings $s,x$  with probability given by
\begin{equation}
    \label{eq3PRL}
    D(s,x)= \dfrac{1}{2^{n.(k-1)}}|\langle x|U_s|0\rangle^{\otimes n}|^2.
\end{equation}
Fixing $t=2$ and $\varepsilon$ to an appropriate value, in this case the value of $k$ becomes $k=O(n)$, we will use this value of $k$ throughout this work. The results of \cite{HBS+17,MB17} directly imply (see also \cite{MGDM19}), that the distribution 
\begin{equation}
\label{eq4PRL}
 D:=\{D(s,x)\}   
\end{equation}
satisfies the following anti-concentration property \cite{MB17,HBS+17}
\begin{equation}
\label{eq5PRL}
    Pr_{s,x}\left(D(s,x) \geq \dfrac{\alpha}{2^{k.n}}\right) \geq \beta,
\end{equation}
where $\alpha$ is a positive constant, $0<\beta \leq 1$, and $Pr_{s,x}(.)$ is the probability over the uniform choice of bit strings $s$ and $x$.

By using the same techniques as \cite{BMS16PRL,HBS+17, MGDM19}, the following proposition can be shown.

\begin{proposition} 
\label{prop1PRL}
Given that the polynomial Hierarchy (PH) does not collapse to its 3rd level, and that the worst-case hardness of approximating the probabilities of $D$ (Equations (\ref{eq3PRL}) and (\ref{eq4PRL})) extends to average-case; there exists a positive constant $\mu$ such that no $poly(n)$-time classical algorithm $C$ exists that can sample from a probability distribution $D_C$  such that
\begin{equation} 
\label{eq6PRL}
\sum_{s,x}|D_C(s,x)-D(s,x)| \leq \mu.
\end{equation}
\end{proposition} 

 Indeed, as shown in \cite{MGDM18}, (\ref{eq2PRL}) can be viewed as implementing a 1D random circuit, as those in \cite{BHH16}. 
 In this picture the circuits have depth $O(n)$ (for fixed $t$ and $\varepsilon$) and are composed of 2-qubit gates which are universal on $U(4)$.
 These circuits are therefore universal under post-selection implying that there exist probabilities $D(s,x)$ which are hard ($\sharp$ P) to approximate up to relative error 1/4+O(1) \cite{FT17} (this property is referred to as \emph{worst-case hardness} of approximating the probabilities of $D$, or for simplicity worst-case hardness).  Worst-case hardness together with the anti-concentration property of Equation (\ref{eq5PRL}) mean that the techniques of \cite{BMS16PRL} directly prove Proposition \ref{prop1PRL}. 
 
 Note that Proposition \ref{prop1PRL} is a conditional statement, meaning that it is true up to some conjectures being true. 
 The first is that the $PH$ does not collapse to its 3rd level, a generalization of $P \neq NP$, which is widely held to be true \cite{gasarch2012guest}. The second conjecture is that the worst-case hardness of the problem extends to average-case, meaning roughly that $most$ outputs are hard to approximate up to relative error 1/4 + O(1). Although this conjecture is less-widely accepted, there exists  evidence to support it  mainly in the case of random circuits sampling unitaries from the Haar measure \cite{bouland2018quantum,movassagh2019cayley}. Particularly relevant to our case are arguments in \cite{BHS+17,MGDM19} which give convincing evidence that worst-case hardness should  extend to average-case for distributions of the form $D(s,x)$ (Equation (\ref{eq3PRL})), where the uniform distribution over bit-strings $s$ effectively makes $D(s,x)$ \emph{more flat} as compared to, say, the outputs of random quantum circuits \cite{bouland2018quantum,movassagh2019cayley} or standard IQP circuits \cite{BMS16PRL}.  Also, in \cite{gao17} an average-case hardness conjecture was stated involving an MBQC construction with fixed $XY$  angles, as is the case here.  Furthermore, we note that a worst-to-average-case conjecture is effectively always required  in  all known proofs of hardness of approximate classical sampling  up to a constant error in the $l_1$-norm \cite{HM17}.
 

\bigskip

The circuit implementing this construction is constant depth.
To see this, notice that the regularly structured graph states of \cite{BHS+17,gao17,HBS+17,HHB+19,MGDM18,MGDM19} can be constructed from constant depth quantum circuits composed of Hadamard ($H$) and $CZ$ gates \cite{HDE+06}. The measurements, being non-adaptive, may be performed simultaneously (depth one).
The explicit form of the circuit can be seen by re-writing the state $|G'\rangle$ as follows
\begin{eqnarray}
\label{EQN: G' as circuit on Ts and 0s}
|G'\rangle = \bigotimes_{a \in V}H_a  \bigotimes_{b\in V_2} && Z_b  \left(\pi/2\right) \prod_{\{i,j\} \in E} CZ_{ij}\nonumber\\
&&\bigotimes_{c\in V_1} |T\rangle_c \bigotimes_{d \in V/V_1} H_d |0\rangle_d,
\end{eqnarray}
where $|T\rangle = Z(\pi/4)H|0\rangle =(|0\rangle + e^{i\pi/4}|1\rangle)/\sqrt{2}$ is referred to as the $T$-state or magic state. Taking out the $T$-state explicitly as here will be useful for applying fault-tolerant techniques.
In this way, these architectures can be viewed as constant depth 2D circuits with NN two-qubit gates
\footnote{A recent paper \cite{NLD+20} shows that the outputs of 2D constant depth circuits are generally efficiently simulable classically. However, we note that the circuits discussed here \cite{BHS+17,gao17,HBS+17,HHB+19,MGDM18,MGDM19} correspond to worst-case instances of the circuits in \cite{NLD+20}, where their efficient classical algorithm fails. Indeed, the $XY$ measurement angles performed effectively induce a 1D dynamics which is purely unitary, and which for the choice of $XY$ angles made in \cite{MGDM18,MGDM19} and here in our case typically evolves an input state onto a \emph{volume law entangled} state. The classical algorithm in \cite{NLD+20} is generally inefficient in simulating such volume law entangled states.}. 

We will show that this constant depth property prevails in our fault-tolerant version of these architectures as well, in our case using 4D and 3D circuits with NN two-qubit gates. As a final remark, note that the 2D NN circuit presented here has the single-instance hardness property, because the choice of measurement angles is fixed \cite{gao17}.


\bigskip

\textbf{\emph{Noise model}} - Before going into details of the fault-tolerant techniques, we present the noise model which we adopt. We will consider the \emph{local stochastic quantum noise} model, following \cite{FGL18,BGK+19}. Local stochastic noise can be thought of as a type of noise where the probability of the error $E$ occuring decays exponentially with the size of its support. This noise model encompasses errors that can occur in qubit preparations, gate applications, as well as measurements. It also allows for the errors between successive time steps of the circuit to be correlated \cite{FGL18,BGK+19}. 
More precisely, following the notation in \cite{FGL18,BGK+19}, a local stochastic noise with rate $p$, where $p$ is constant satisfying $0<p<1$, is an $m$-qubit Pauli operator $$E=\otimes_{i=1,...,m}P_{i},$$ where $P_{i} \in \{1,X,Y,Z\}$ are the single qubit Pauli operators, such that 
$$Pr\big( F \subseteq Supp(E)\big) \leq p^{|F|},$$
for all $F \subseteq \{1,...,m\}$, where $Supp(E) \subseteq \{1,..,m\}$ is the subset of qubits for which $P_{i} \neq 1$. Also following notation in \cite{FGL18,BGK+19}, we will denote a local stochastic noise with  rate $0<p<1$ as  $E \sim \mathcal{N}(p)$.

We will use the following property of local stochastic noise, shown in \cite{BGK+19}, which says that all errors for constant depth Clifford circuits can be pushed to the end. Consider a constant depth-$d$ noiseless quantum circuit \begin{equation*}
U=U_{d}...U_{1}, 
\end{equation*}
which acts on a prepared input state and is followed by measurements, where each $U_{i}$ for $i=\{1,...,d\}$ is a depth-one circuit composed of single and two-qubit Clifford gates. It was shown in \cite{BGK+19} that a noisy version of this circuit satisfies
\begin{eqnarray}
\label{eq7PRL}
U_{noisy}&=& E_{out}.E_{d}U_{d}....E_{1}U_{1}E_{prep} \nonumber \\ \nonumber
&=& E(U_d...U_1) \\
&=& EU,
\end{eqnarray}
where $E_i \sim \mathcal{N}(p_i)$ for $i \in \{1,...,d\}$, with constant $0<p_{i}<1$ is the noisy implementation of depth-one circuit $U_i$, $E_{prep} \sim \mathcal{N}(p_{prep})$ and $E_{out} \sim \mathcal{N}(p_{out})$ with constants $0<p_{prep},p_{out}<1$ are the errors in the preparation and measurement respectively \footnote{Note that by choosing different values of $p_{prep}$, $p_{out}$, and $p_i$ one can differentiate between the noises of preparation, gate application, and measurement. One can also account for scenarios where some operations could be more faulty than others, as is commonly done for example when assuming two-qubit gates are faultier than single qubit gates \cite{Li15}. }.

For constant depth $d$, $E \sim \mathcal{N}(q)$ where $0<q<1$ is a constant which is a function of $p_1,...,p_d,p_{prep},p_{out}$ \cite{BGK+19} \footnote{For example, when  $p_{prep}=p_{out}=p_{1}=...=p_{d}=p$, then $q \leq p^{4^{-d-1}}$ \cite{BGK+19}. Note that $p^{4^{-d-1}}$ is a constant when $d$ is a constant, meaning that $q$ is upper-bounded by a non-zero constant. For a suitable choice of $p$ we can therefore tune $q$ to be below the threshold of fault-tolerant computing with the surface code, where the classical decoding fails with a probability decaying exponentially with the code distance \cite{BGK+19}.  }. Equation (\ref{eq7PRL}) shows that the errors accumulating in a constant depth quantum circuit composed of single and two qubit Clifford gates can be treated as a single error $E$. Furthermore, for small enough $q$ (i.e small enough $p_1,...,p_d,p_{prep},p_{out}$ - typically, these should be smaller than the threshold of fault-tolerant computing with the surface code \cite{BGK+19,DKL+02}  or of the 3D cluster state \cite{RHG07} in our case), $E$ can be corrected with high probability by using standard techniques in quantum error correction (QEC) \cite{BGK+19,wang2011surface}. Also, $E$ can be propagated until after the measurements, where the error correction procedure is completely classical. 

\bigskip
\section*{4D NN architecture}
In this part of the paper, we will describe the construction of our 4D NN architecture demonstrating a quantum speedup. Our approach takes three ingredients, the sampling based on regular graph states mentioned above \cite{gao17,BHS+17,HBS+17,MGDM18,MGDM19,HHB+19}, fault-tolerant single shot preparations of logical qubit states \cite{BGK+19}, and magic state distillation (MSD) \cite{BK05,HHP+17,Li15}.
A large part of fault-tolerant techniques follow the work of \cite{BGK+19}, where they present a family of constant depth circuits which give statistics that cannot be reproduced by any classical computer of constant depth. To do so they introduce error correcting codes where it is possible to prepare logical states fault-tolerantly with constant depth, and Clifford gates are transversal. Then, they also show that for local stochastic quantum noise, all errors for Clifford circuits can be traced through to effectively be treated as a final error, meaning that errors do not have to be corrected during the circuit. Together these allow for constant depth fault-tolerant versions of constant depth Clifford circuits. Compared to \cite{BGK+19}, the big difference in our work is the need for non-Clifford operations (for the choice of local measurement angle). To address this, we use so called magic states which can be distilled fault-tolerantly \cite{BK05}. Generally their distillation circuits  are not constant depth however, and here we adapt the distillation  circuits of \cite{HHP+17} to be constant depth using ideas from MBQC. 
In particular we do not use feed-forward in the distillation procedure, and instead translate depth  of circuits for cost of having to do many copies of constant depth circuits (each being an MSD circuit with no feed-forward) many times in parrallel. 
We show that, for specific MSD techniques \cite{HHP+17,HH18,Jones13,HH182}, a balance can be reached which gives sufficiently many magic states of high enough fidelity to demonstrate quantum speedup in constant depth with polynomial overhead in number of ancillas.
We then use MBQC notions to route in the high fidelity magic states into our sampling circuit. This is also done in constant depth.  At this point, interaction with a classical computer is required. This is mainly in order to identify which copies of MSD circuits (which are done in parallel) were successful in distilling magic states of sufficiently high fidelity. After, these high fidelity magic states are taken, together with more ancillas, to make a logical version of the graph state, which is then measured. Effectively we then have two constant depth quantum circuits with an efficient (polynomial) classical computation in between.

The constant depth MBQC distillation, together with the constant depth MBQC routing  will ensure that enough magic states with adequately high fidelity are  always injected into our sampling problem, thereby  enabling us to observe quantum speedup \emph{deterministically} at each run of the experiment, since we would determinstically recover an encoded version of the 2D NN architecture with the single-instance hardness property described in earlier sections \cite{MGDM18}. This is contrary to what happens in \cite{KT19}, where an encoded version of this 2D NN architecture is constructed \emph{probabilistically}, albeit with exponentially low probability of success.
\bigskip

\textbf{\emph{Logical encoding}} - 
Following \cite{BGK+19}, we use the folded surface code \cite{bravyi1998quantum,Moussa16,BGK+19}. A single logical qubit is encoded into $l$ physical qubits. We denote the logical versions of states and fault-tolerant gates using a bar, that is, a state $|\psi\rangle$ of $m$ qubits would be encoded onto its logical version $|\overline{\psi}\rangle$ on $m.l$ qubits and operator $U$ would be replaced by logical operator $\overline{U}$.
The choice of encoding onto the folded surface code has two main advantages, firstly, Clifford gates have transversal fault-tolerant versions, meaning the fault-tolerant versions of a constant depth Clifford circuit are also constant depth and composed of  single and two-qubit Clifford gates acting on physical qubits of the code \cite{BGK+19}. For example $$\overline{X}=\bigotimes_{i \in V_{diag}} X_i,$$
 where $V_{diag}$ is the set of physical qubits lying on the main diagonal of the surface code, $X_i$ is a Pauli $X$ operator acting on physical qubit $i$. Similarly for the logical version of the Pauli $Z$ operator
 $$\overline{Z}=\bigotimes_{i \in V_{diag}} Z_i.$$
  Secondly, the preparation of the logical $|\overline{0}\rangle$ and $|\overline{T}\rangle$ states can be done fault-tolerantly in constant depth \cite{Li15,BGK+19}.

\begin{figure*}
\begin{center}
\graphicspath{}
\includegraphics[trim={2 0cm 0 1cm} , scale=0.5]{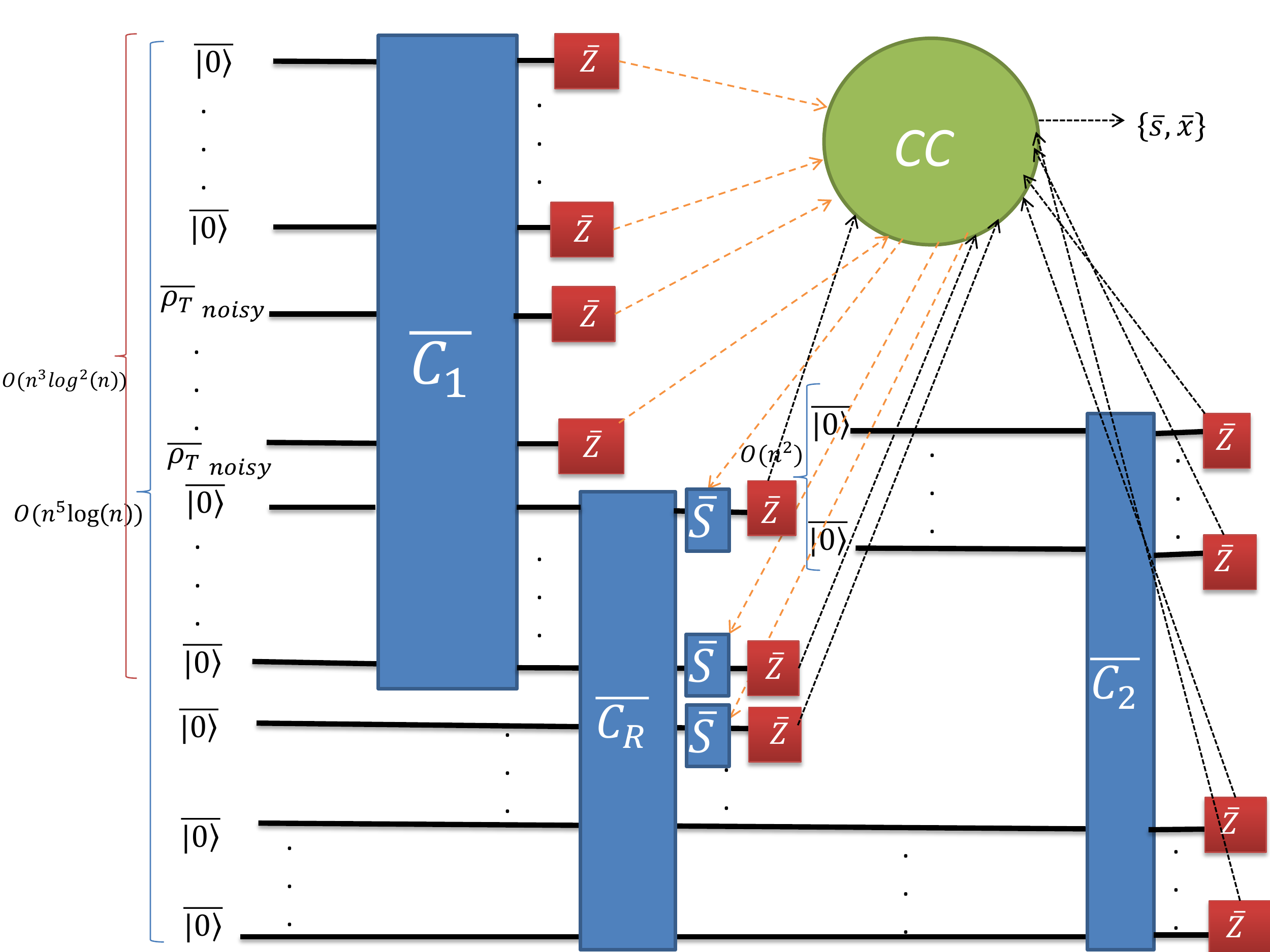}
\caption{Overview of the 4D NN circuit for our sampling problem. The overall circuit takes in noisy logical $|\overline{0}\rangle$ 
and $\overline{\rho_T}_{noisy}$ states, which can be prepared in constant depth \cite{Li15,BGK+19} (up to Paulis, which can be traced through and dealt with efficiently classicaly after measurements of the logical qubits of our circuit, as described in the main text).
It is composed of three underlying circuits, $\overline{C_1}$, which implements the MSD, then $\overline{C_R}$ 
which routes the good outputs to the final circuit $\overline{C_2}$ which generates the graph state $|\overline{G^{'}\rangle}$. 
The construction also calls on a classical computer ($CC)$ to process correction operations, indicated by the dotted lines of different colours. 
The orange dotted lines are in order to identify the successful MSD outputs, and create the paths for routing them. 
The $\overline{S}$ gate is either a $\overline{H}$ gate or an identity gate, depending on the classical control.
The black dotted lines are the  measurement results of non-output qubits of $\overline{C_R}$, and the measurement results of qubits of $\overline{C_2}$ which are fed into the classical computer which performs a postprocessing to output the final sample $\{\overline{s},\overline{x}\}$ (Equation (\ref{eq14PRL})). }
\label{FIG Overview circuit}
\end{center}
\end{figure*}

 The preparation of the logical $|\overline{0}\rangle$ state can be done fault-tolerantly using the single-shot preparation procedure of \cite{BGK+19}. This requires a constant depth 3D quantum circuit, together with polynomial time classical post-processing, which can be pushed until after measurements of logical qubits of our circuit (see Figure \ref{FIG Overview circuit}).
 This constant depth quantum circuit consists of non-adaptive measurements on a 3D cluster state composed of $O(l^{3/2})$ (physical) qubits \cite{BGK+19}. 
 The 3D cluster state being of regular structure can be prepared in constant depth. The non-adaptive measurements create a two-logical qubit Bell state up to a Pauli operator. 
 The classical post-processing is in order to trace these Paulis through the Clifford circuits (Figure \ref{FIG Overview circuit}) and correct the measurement results accordingly.
 In \cite{BGK+19} it is shown that this preparation process is fault-tolerant, by showing that, in the presence of local stochastic quantum noise the overall noise induced from the preparation, measurements, and Pauli correction is a local stochastic noise with constant rate \cite{BGK+19}.
 For our purposes, we will only use one logical qubit of the Bell state 
 \footnote{One way to do this would be measuring the other logical qubit of the Bell state non-adaptively in $\overline{Z}$, then decoding the result and applying an $\overline{X}$ to the unmeasured logical qubit dependant on the decoded measurement result. This should be done after the recovery Pauli operator of \cite{BGK+19} has been applied. The noise acting on the unmeasured qubit after completion would still be local stochastic with constant rate. Indeed, after applying the recovery operator of \cite{BGK+19}, we are left with a Bell state with some local stochastic noise $E$ \cite{BGK+19}, then after measuring one logical qubit and decoding (which succeeds with high probability if error rates are small), we apply a conditional $\overline{X}$ operator to the unmeasured logical qubit. In the case this $\overline{X}$ is applied, it introduces also a local stochastic noise $E^{'}$, but because $\overline{X}$ is a constant depth Clifford gate with only single qubit gates, $E^{'}$ can be merged with  $E$ to give a single local stochastic noise $E^{''}$ which is still local stochastic with constant rate, by the likes of arguments of Equation (\ref{eq7PRL}). In what remains, we incorporate this operation into the classical post-processing needed to apply the recovery operator of the single-shot preparation procedure of \cite{BGK+19}. We will therefore mean by recovery operator hereafter, the Pauli recovery operator of \cite{BGK+19} together with the conditional $\overline{X}$ which is applied to the unmeasured logical qubit. Note also that, as mentioned in the main text, we will often push applying this recovery operator until later parts of the circuit (for example after measuring the non-outputs of all copies of $zMSD$ as well as after the final measurements of $\overline{C_2}$), in that case the arguments for the overall noise being local stochastic still hold and follow similar reasoning as above. }. 
 
The preparation of the logical $T$-state  $|\overline{T}\rangle$ can also be done in constant depth by using a technique similar to \cite{Li15}. 
 Indeed, in the absence of noise, a perfect logical $T$-state can be prepared by the initialization of $l$ physical  qubits  (over a constant number of rounds), as well as three rounds of full syndrome measurements; as detailed in \cite{Li15} 
 \footnote{ The constant depth procedure of \cite{Li15} also requires some post-selection (in the presence of noise). 
 However, this post-selection is usually over measurement results of a small (constant) number of qubits, and the success probability is also a constant \cite{Li15}. We can therefore implement in paralell $O(1)$ runs of this constant depth procedure, and we are guaranteed with high probability that at least one run corresponds to the desired post-selection. }. 
 Each of the syndrome measurement rounds, because of the locality of the stabilizers in the surface code, can be scheduled in such a way as to be implemented by a constant depth quantum circuit composed of Controlled Nots and ancilla qubit measurements \cite{LAR+11,Li15}. In the presence of noise, this procedure prepares a noisy logical $T$-state (Equation (\ref{eqnoisyy})), starting from a noisy physical qubit $T$-state, and noisy preparations, gates and measurements \cite{Li15} 
 \footnote{ Although the noise model used in \cite{Li15} is not the same as the one we use here, where in \cite{Li15} they use independent depolarizing noise for preparations and gate application, and with different rates for single and two-qubit gates, we believe their results hold in our case as well. Indeed, viewing a local stochastic noise with rate $p$ on a single qubit, this qubit could experience an error (after preparation, measurement or gate application) with probability $pr \leq p$ (from the definition of local stochastic noise with $|F|=1$, see main text), this is in line with the noise model of \cite{Li15} where the probability of error is exactly $p$. Furthermore, choosing different error rates for local stochastic noise applied after single and two-qubit gates allows mimicking what happens in the noise model of \cite{Li15}. }. However, distillation is required to get sufficiently high quality $T$-states, which will be dealt with separately later. For simplicity, for now we will assume perfect $T$-states.

\bigskip

Starting with the prepared logical $|\overline{0}\rangle$ and $|\overline{T}\rangle$, the logical version of Equation (\ref{EQN: G' as circuit on Ts and 0s}) is written in terms of the constant depth circuit $\overline{C_2}$,
\begin{eqnarray}
|\overline{G'}\rangle = \overline{C_2} \bigotimes_{c\in V_1} |\overline{T}\rangle_c \bigotimes_{d \in V/V_1} |\overline{0}\rangle_d
\label{EQN: logical G'}
\end{eqnarray}
where 
\begin{equation}
    \overline{C_2} := \bigotimes_{a \in V}\overline{H_a}  \bigotimes_{b\in V_2} \overline{Z_b}  \left(\pi/2\right) \prod_{\{i,j\} \in E} \overline{CZ_{ij}} \bigotimes_{d \in V/V_1} \overline{H_d}.
    \label{EQN: C_2}
\end{equation}
Since all gates are Clifford, the physical circuit implementing $\overline{C_2}$ is constant depth. This circuit is the last circuit element in Figure~\ref{FIG Overview circuit} which combines the elements of our construction.

The logical $\overline{Z}$  measurements are carried out by physical $Z$ measurements on the physical qubits of the surface code, and 
 several classical  decoding algorithms have been established \cite{BGK+19,wang2011surface,LAR+11,DKL+02}. 
 In the noiseless case, the decoding algorithm consists of calculating the sum (modulo 2) of the measurement result from measuring $Z$ on the physical qubits of the main diagonal of the surface code. 
 In the presence of noise, the decoding algorithm takes as input the (noisy) measurement results of all the $l$ physical qubits of the surface code which are measured in the same basis as the qubits on the main diagonal, and performs a minimal weight perfect matching to correct for the error induced by the noise \cite{BGK+19,fowler2012proof,DKL+02}. 
 For small enough error rates (below the threshold of fault-tolerant computing with the surface code), the probability that these decoding algorithms fail, that is, the probability that the noise changes the parity of the $\overline{Z}$ measurement result after decoding, decreases exponentially with the code distance $cd$, which for surface codes scales as $cd=O(\sqrt{l})$ \cite{bravyi1998quantum,BGK+19,DKL+02,fowler2012proof}. 

Let $$\overline{s}=\{\overline{s}_1,...,\overline{s}_{n.(k-1)}\},$$ denote the measurement results of the logical qubits of all but the last column of $|\overline{G'}\rangle$. Similarly, let 
 $$\overline{x}=\{ \overline{x}_1,...,\overline{x}_n\},$$ denote the measurement results of the logical qubits of the last column of $|\overline{G'}\rangle$.
 
If we call $\overline{D}(\overline{s},\overline{x})$ the probability of getting $(\overline{s},\overline{x})$ in the absence of noise, it follows straightforwardly from the logical encoding that
 \begin{equation}
 \label{eq9PRL}
     \overline{D}(\overline{s},\overline{x})=D(s,x),
 \end{equation}
for all $\overline{s} \in \{0,1\}^{n.(k-1)}$, and $\overline{x} \in \{0,1\}^{n}$, where $D(s,x)$ is as defined in Equations (\ref{eq3PRL}) and   (\ref{eq4PRL}). 
That is, in the absence of noise, measuring non-adaptively the logical qubits of $|\overline{G'}\rangle$ in  $\overline{Z}$ defines a sampling problem with probability distribution $\overline{D}$ demonstrating a quantum speedup, by Proposition \ref{prop1PRL}.

We will now see that this sampling remains robust under local stochastic noise. 
Noise must be addressed at each part of the construction. 
The first being that each depth-one step of the circuit preparing $|\overline{G'}\rangle$ is now followed by a local stochastic noise, as in the example of Equation (\ref{eq7PRL}). 
Also, the single-shot preparation procedure of \cite{BGK+19} becomes noisy, however as shown in \cite{BGK+19} this noise is local stochastic with constant error rate and therefore can be treated as a preparation noise in preparing $|\overline{G'}\rangle$, analogous to $E_{prep}$ in Equation (\ref{eq7PRL}). As seen earlier, the circuit preparing $|\overline{G'}\rangle$ is constant depth and composed of single and two-qubit Clifford gates acting on physical qubits. 
Therefore, we can use the result of \cite{BGK+19}, which is shown in Equation (\ref{eq7PRL}), and treat all the noise accumulating through different steps of the circuit as a single local stochastic noise $E \sim \mathcal{N}(q)$ with a constant rate $0<q<1$, acting on the (classical) measurement outcomes \cite{BGK+19}. Therefore, when $q$ is low enough \cite{BGK+19}, $E$ can be corrected with high probability using the classical decoding algorithms described earlier \cite{BGK+19}.  
In appendix \ref{APPC}, we show that when the number of physical qubits per logical qubit $l$ scales as
\begin{equation}
\label{eq10PRL}
    l \geq O(log^2(n)),
\end{equation}
where $n$ is the number of rows of $|G\rangle$ \footnote{ this is usually the input part of an MBQC \cite{RB01}, which is the basis of our construction }, this suffices for our needs.

More precisely, we denote $\tilde{\overline{D}}_1(\overline{s},\overline{x})$ the probability of getting outcomes $(\overline{s},\overline{x})$ in the presence of stochastic noise, after performing a classical decoding of the measurement results \cite{BGK+19}, but where logical $T$-states are assumed perfect (noisless). 
Then, if $l$ satisfies Equation (\ref{eq10PRL}), and for small enough  error rates (below the threshold of fault-tolerant computing with the surface code \cite{DKL+02}) of preparations, single and two-qubit gates, as well as measurements, $\tilde{\overline{D}}_1(\overline{s},\overline{x})$ can be made $1/poly(n)$ close in $l_1$-norm to the noiseless version  (Equation (\ref{eq9PRL})). That is,
\begin{equation}
    \label{eq11PRL}
    \sum_{\overline{s},\overline{x}}|\tilde{\overline{D}}_1(\overline{s},\overline{x})-\overline{D}(\overline{s},\overline{x})| \leq \dfrac{1}{poly(n)}.
\end{equation}

This means that for a given constant $\mu_1$, there exists a large enough constant $n_0$, such that for all $n \geq n_0$ classically sampling from $\tilde{\overline{D}}_1$  up to $l_1$-norm error $\mu-\mu_1$  implies, by a triangle inequality, sampling from $\overline{D}$ up to $l_1$-norm error $\mu$, which presents a quantum speedup by Proposition \ref{prop1PRL} \cite{BMS16PRL}. Therefore, we have recovered  quantum speedup in the presence of local stochastic noise, assuming perfect $T$-states.

\bigskip

\textbf{\emph{Distillation of $T$-states}}- The final ingredient is the distillation of the $T$-states.
The analysis we have done so far assumes we can still prepare perfect logical $T$-states. In reality, however, this is not the case. Indeed, in the presence of noise, the constant depth preparation procedure of \cite{Li15} can only prepare a logical $T$-state with error rate $0<\varepsilon<1$
\begin{equation}
\label{eqnoisyy}
    \overline{\rho_T}_{noisy}:=(1-\varepsilon)|\overline{T}\rangle \langle \overline{T}| + \varepsilon \eta,
\end{equation}
with $\eta$ an arbitrary $l$-qubit state. In order to get high purity logical $T$-states, one must employ a technique called magic state distillation (MSD) \cite{BK05}. An MSD circuit is a Clifford circuit which usually takes as input multiple copies of noisy $T$-states $\overline{\rho_T}_{noisy}$, together with some ancillas, and involves measurements and post-selection in order to purify these noisy input states \cite{BK05}. The output of an MSD circuit is a logical $T$-state $\overline{\rho_T}_{out}$  with higher purity than the input one. That is, 
\begin{equation}
    \label{eqPRL12}
    \overline{\rho_T}_{out}:=(1-\varepsilon_{out})|\overline{T}\rangle\langle\overline{T}| + \varepsilon _{out}\eta^{'},
\end{equation}
with $0<\varepsilon_{out}<\varepsilon<1$, and $\eta^{'}$ an arbitrary $l$-qubit state. For small enough $\varepsilon$ \cite{reichardt2005quantum} \footnote{ This is guaranteed by using the technique of \cite{Li15} if the error rate of preparations, single and two-qubit gates is low enough. Since $\varepsilon$ in \cite{Li15} is generally a function of these error rates. }, $\varepsilon_{out}$ could be made arbitrarily small by repeating the MSD circuit an appropriate number of times \cite{BK05}. 

MSD circuits need not in general be constant depth. 
Our approach to depth is, again, via a translation to the measurement based quantum computing (MBQC) paradigm \cite{RB01}.
In MBQC one starts off with graph state, for example the 2D grid cluster state, and computation is carried out through consecutive measurements on individual qubits. 
In order to preserve determinism these measurements must be corrected for. 
For a general computation this must be done round by round (the number of rounds typically scales with the depth of the corresponding circuit, though there can be some separation thereof \cite{browne2007generalized}). 
If we forgo these corrections, we end up applying different unitaries, depending on the outcome of the measurement results - indeed, this is effectively what happens in Equation (\ref{eq2PRL}).
Thinking of MBQC now as a circuit, if one could do all measurements at the same time, one could think of it as a constant depth circuit, since all that is needed is to construct the 2D cluster state followed by one round of measurements and corrections, which can be done in constant depth. 
This is possible for circuits constructed fully of Clifford operations, but not generally, and not for the MSD circuits we use here because of the $T$ gates (or feedforward), so we are forced to sacrifice determinism.

Now, in order to get constant depth MSD, we translate the MSD circuits in \cite{HHP+17} to MBQC. The choice of this MSD construction is argued in appendix \ref{APP subsec zMSD}.
Since we want to maintain constant depth, we want to perform all measurements at the same time, however the cost is that it will only succeed if we get the measurement outcomes corresponding to the original circuit of \cite{HHP+17} with successful syndromes.
In order to produce enough $T$ states, the trick is simply to do it many times in parallel. 
That is, we will effectively implement many copies of the MBQC computation, so that we get enough successes. Effectively we trade depth of the corresponding circuit for number of copies and ancillas. Fortunately, for our specifically chosen MSD protocols \cite{HHP+17,HH182}, we will see that this cost is not too high.

Furthermore, this is all done in the logical encoding of the folded surface code. Our construction for this, which we denote $zMSD$, is designed to take copies of the noisy encoded $T$-states (\cite{Li15}) and ancilla in the encoded $|\overline{0}\rangle$ state, and affect $z$ iterations of the fault-tolerant version of MSD protocol in \cite{HHP+17}. As discussed above, this happens only when the correct results occur in the MBQC. In this case we say the $zMSD$ was \emph{successful}.
We denote the circuit version of this as $\overline{C_1}$ (see Figure \ref{FIG Overview circuit}). 
In appendix \ref{app zMSD}, we show that when $zMSD$ is successful, $\varepsilon_{out}$ satisfies
\begin{equation}
\label{eq13PRL}
    \varepsilon_{out} \leq O(\dfrac{1}{n^4}).
\end{equation}
We also show that performing $O(n^3log(n))$ copies of $zMSD$ circuits (which can be done in parallel), each of which is composed of $O(log(n))$ logical qubits as seen in appendix \ref{APP subsec zMSD}, guarantees with high probability
\begin{equation}
\label{eqpsuccprl}
    p_{succ} \geq 1-\dfrac{1}{e^{poly(n)}},
\end{equation}
that at least $O(n^2)$ copies  of $zMSD$ will be successful (we will refer to these often as successful instances of $zMSD$). Note that $O(n^2)=O(k.n)$ is the number of perfect logical $T$-states needed to create $|\overline{G'}\rangle$ \cite{MGDM18}. Furthermore, because $zMSD$ is constant depth and composed of single and two-qubit Clifford gates, errors can be treated as a single local stochastic noise after the measurements with constant rate (see Equation (\ref{eq7PRL})) which can be corrected classically with high probability when the error rates are low enough using the standard decoding algorithms described previously \cite{fowler2012proof,BGK+19}.

The remaining task is to route these good states into the inputs of the circuit $\overline{C_2}$ (Equation (\ref{EQN: C_2}))
depending on the measurement outcomes - i.e. make sure that only the good outputs go to make $|\overline{G'}\rangle$. The most obvious approach, using control SWAP gates, results in a circuit whose depth scales with $n$. Here, once more, we use MBQC techniques in order to bypass additional circuit depth. The idea is to feed the outputs through a $2D$ cluster graph state, and dependent on the measurement results of the $zMSD$, the routing can be etched out by Pauli Z measurements. Since the graph is regular, and, since the measurements can be made at the same time, this can be done in constant depth, up to Pauli corrections (which can be efficiently traced and dealt with by the classical computation at the end).
We denote the fault-tolerant circuit implementing this as $\overline{C_R}$, see Figure \ref{FIG Overview circuit}.
Details of the construction can be found in appendix \ref{APP routing}, where we also show that errors remain manageable.

Finally, we denote
$\tilde{\overline{D}}_2(\overline{s},\overline{x})$  to mean the probability of observing the outcome $(\overline{s},\overline{x})$ after measuring all logical qubits after $\overline{C_2}$ (Equation (\ref{EQN: C_2})), in the presence of local stochastic noise, and where each $T$-state fed into $\overline{C_2}$ is replaced by  $\overline{\rho_T}_{out}$, and performing a classical decoding of these measurement results \cite{BGK+19}.
Then, we show, in appendix \ref{app zMSD}, that when $\varepsilon_{out}$ satisfies Equation (\ref{eq13PRL}), 
\begin{equation}
    \label{eq14PRL}
    \sum_{\overline{s},\overline{x}}|\tilde{\overline{D}}_2(\overline{s},\overline{x})-\overline{D}(\overline{s},\overline{x})| \leq \dfrac{1}{poly(n)}.
\end{equation}
Therefore, by the same reasoning as that for $\tilde{\overline{D}}_1$, for small enough error rates, for large enough $n$, and with very high probability $p_{succ}$, we can prepare a constant depth quantum circuit sampling from a noisy distribution $\tilde{\overline{D}}_2$ under local stochastic noise, presenting a quantum speedup.

 Our main result can therefore be summarized in the following Theorem, whose proof follows directly from showing that Equation (\ref{eq14PRL}) holds and using Proposition \ref{prop1PRL}. 

\begin{theorem}
\label{TH1}
Assuming that the $PH$ does not collapse to its third level, and that worst-case hardness of the sampling problem  (\ref{eq4PRL}) extends to average-case.
There exists a positive constant $0<p<1$, and a positive  integer $n_o$, such that for all $n \geq n_o$, if the error rates of local stochastic noise in all preparations, gate applications, and measurements in $\overline{C_1}$, $\overline{C_R}$, and $\overline{C_2}$ are upper-bounded by  $p$, then with high probability $p_{succ}$ (Equation (\ref{eqpsuccprl})), the sampling problem  $\tilde{\overline{D}}_2$  defined by (\ref{eq14PRL}) can be constructed, and  no $poly(n)$-time classical  algorithm exists which can sample from $\tilde{\overline{D}}_2$  up to a constant $\mu^{'}$ in $l_1-$norm .
\end{theorem}
\bigskip

\textbf{\emph{Overview of the 4D NN architecture}} - The overall construction is presented in Figure \ref{FIG Overview circuit} as a combination of the three circuits mentioned above, $\overline{C_1}$ implementing the MSD, the routing of successful $T$-states in $\overline{C_R}$, and the circuit for the construction of the state $|\overline{G'}\rangle$ in $\overline{C_2}$. Overall it takes the noisy logical $|\overline{0}\rangle$ and $\overline{\rho_T}_{noisy}$ states as inputs and the final measurements are fed back to a classical computer ($CC$) to output the error corrected results $\overline{s},\overline{x}$, according to distribution $\tilde{\overline{D}}_2$ (Equation (\ref{eq14PRL})). The preparation of the logical input states is done in constant depth \cite{BGK+19,Li15} and each of these three composite circuits are constant depth, using at most three dimensions.  Furthermore,  assuming that classical computation is instantaneous, our entire construction can be viewed as a constant depth quantum circuit. Indeed, as already seen $\overline{C_2}$ is constant depth, what remains is to show the same for $\overline{C_1}$ and $\overline{C_R}$. We show this in appendix \ref{APP subsec zMSD} and \ref{APP routing}. 


During the circuit, we require some side classical computation, which inputs back into the circuit at one point. Classical information to and from the classical computer are indicated by dotted orange and black lines in Figure \ref{FIG Overview circuit}.
First, the measurements of the non-outputs for the $zMSD$ in $\overline{C_1}$,  along with measurement results (not illustrated in figure) of (physical) qubits used in preparing $|\overline{0}\rangle$ \cite{BGK+19} states  making up the copies of $zMSD$, are fed  into the classical computer in order to determine the choice of measurements after the routing circuit $\overline{C_R}$, as indicated by the orange dotted lines in Figure \ref{FIG Overview circuit}.
This part simply identifies the successful $zMSD$ outcomes, followed by calculating the routing path. 
This is the only point that classical results are fed back into the circuit, all other classical computations can be done after the final measurements. After these final measurements, the remaining measurements are fed back into the computer, indicated by black dotted lines in Figure \ref{FIG Overview circuit}. Together with the measurement results from the state preparations \cite{BGK+19} (not illustrated in the figure) these are incorporated into the classical error correction \cite{BGK+19,edmonds1973operation} giving  the outputs $\overline{s},\overline{x}$ with probabilities $\tilde{\overline{D}}_2$.
The classical computation can be done in $poly(n)$-time \cite{RBB03,wang2011surface,edmonds1973operation}.


The total number of physical qubits required scales as  $O(n^5poly(log(n)))$ (where $n$ scales the size of the original sampling problem (Proposition \ref{prop1PRL})). 
This breaks down as follows. 
$\overline{C_1}$ takes as input $O(n^3log^2(n))$ noisy logical $T$-states $\overline{\rho_T}_{noisy}$ and $O(n^3log^2(n))$ ancillas prepared in $|\overline{0}\rangle$. 
$\overline{C_R}$ takes  the outputs of $\overline{C_1}$, and additional $O(n^5log(n))$ logical ancillas prepared in $|\overline{0}\rangle$. This dominates the scaling.
$\overline{C_R}$ sends $O(n^2)$  distilled $T$-states to $\overline{C_2}$, which also takes in $O(n^2)$ copies of $|\overline{0}\rangle$. This means that in total we would need $O(n^5log(n))$ logical qubits. Now, each logical qubit is composed of $l \geq O(log^2(n))$ physical qubits (Equation (\ref{eq10PRL})), and some of these logical qubits ,which need to be prepared in $|\overline{0}\rangle$, require an additional overhead of $O(l^{\frac{3}{2}}) \geq O(log^3(n)))$ physical qubits, as seen previously (see also \cite{BGK+19}).
Therefore, the total number of physical qubits needed is $\sim O(n^5log^4(n)))=O(n^5poly(log(n))).$  

A crucial question relevant to experimental implementations would be calculating the exact values of the error rates of measurements, preparations, and gates needed to achieve fault-tolerant quantum speedup in our construction. Because the quantum depth of our construction is constant and composed of single and two-qubit Clifford gates (as seen previously), we know from \cite{BGK+19} and the likes of Equation (\ref{eq7PRL}) that these error rates are non-zero constants independent of $n$. However, their values may be pessimistically low.  A crude estimate of this error rate is $p \sim e^{-4.6 \times 4^{-d-1}}$. This is assuming preparations (including preparation of noisy logical $T$-states for distillation), measurements, and gates all have the same error rate $p$. $d$ is a constant which is the total quantum depth of our construction, which is the sum of the depths of all preparations, gate applications and measurements involved in constructing $zMSD$, routing the outputs of succesful instances of $zMSD$, and constructing $|\overline{G'}\rangle$. This expression is obtained by using the same techniques as \cite{BGK+19}, where the error rate $q$ of $E$ in Equation (\ref{eq7PRL}) is chosen such that it satisfies $q \leq 0.01$. This is in order for classical decoding to fail with probability decaying exponentially with the code distance of the surface code \cite{BGK+19,fowler2012proof}.

This construction is a constant depth quantum circuit implementable on a 4D NN architecture (or a 3D architecture with long range gates). The reason for this is that our original (non fault-tolerant) construction is a 2D NN architecture \cite{MGDM18} as seen previously, and the process of making this architecture fault-tolerant requires adding an additional two dimensions \cite{BGK+19}, albeit while keeping the quantum depth constant, as explained earlier. If we do not want to use long range transversal $CZ$ gates in 3D, and want all the $CZ$ gates  to be NN, the only way to do this is to work in 4D. Note that this was not a problem in \cite{BGK+19}, as there the original (non fault-tolerant) circuit was a 1D circuit, and introducing fault-tolerance added two additional dimensions, making their construction constant depth with NN gates in 3D \cite{BGK+19}. Nevertheless, we will show in the next section how to make our construction constant depth in 3D with NN two-qubit gates. We will do this by avoiding the use of transversal gates to implement encoded versions of two-qubit gates; a feature which is naturally found in defect-based topological quantum computing \cite{RHG07}. Armed with the ideas of constant depth MSD and MBQC routing, we shall present in this next section a constant quantum depth fault-tolerant construction demonstrating a quantum speedup with only nearest neighbor $CZ$ gates in 3D.
\bigskip

\section*{3D NN architecture}
In this part of the paper, we will explain how the construction for fault-tolerant quantum speedup described earlier can be achieved using a 3D NN architecture, based on the construction of Raussendorf, Harrington, and Goyal (RHG) \cite{RHG07}. Note that in this construction (henceforth referred to as RHG construction), two types of magic states need to be distilled, the $T$-states seen previously, as well as the $Y$-states. A perfect (noiseless) $Y$-state is given by
\begin{equation*}
    |Y\rangle:=\dfrac{1}{\sqrt{2}}(|0\rangle+ e^{i\frac{\pi}{2}}|1\rangle).
\end{equation*}
This state is a resource for the phase gate $Z(\pi/2)$.
The noisy $Y$-state $\rho_{Y_{noisy}}$ is defined analogously to a noisy $T$-state seen earlier
\begin{equation*}
    \rho_{Y_{noisy}}:=(1-\varepsilon)|Y\rangle \langle Y| + \varepsilon \eta,
\end{equation*} 
with $0<\varepsilon<1$ representing the noise, and $\eta$ an arbitrary single qubit state.
 As already mentioned, the RHG construction was also used in \cite{KT19} to achieve fault-tolerant quantum speedup. However, our construction will differ from \cite{KT19} in mainly two ways. The first, as already mentioned, is that our construction deterministically produces a hard instance, whereas that in \cite{KT19} produces such an instance with exponentially low probability.  Secondly, our sampling problem verifies the anti-concentration property by construction \cite{MGDM18}, as explained previously, whereas in \cite{KT19}, this anti-concentration was conjectured. Therefore, in our proofs we assume one less complexity theoretic conjecture ( we use two conjectures in total, see Theorem \ref{TH1} and Proposition \ref{prop1PRL}) as compared to \cite{KT19}. Note that we assume the minimal number of complexity-theoretic conjectures needed to prove quantum speedup, using all currently known techniques \cite{HM17}. 
 
 \begin{figure*}
\begin{center}
\graphicspath{}
\includegraphics[trim={2 0cm 0 1cm} , scale=0.5]{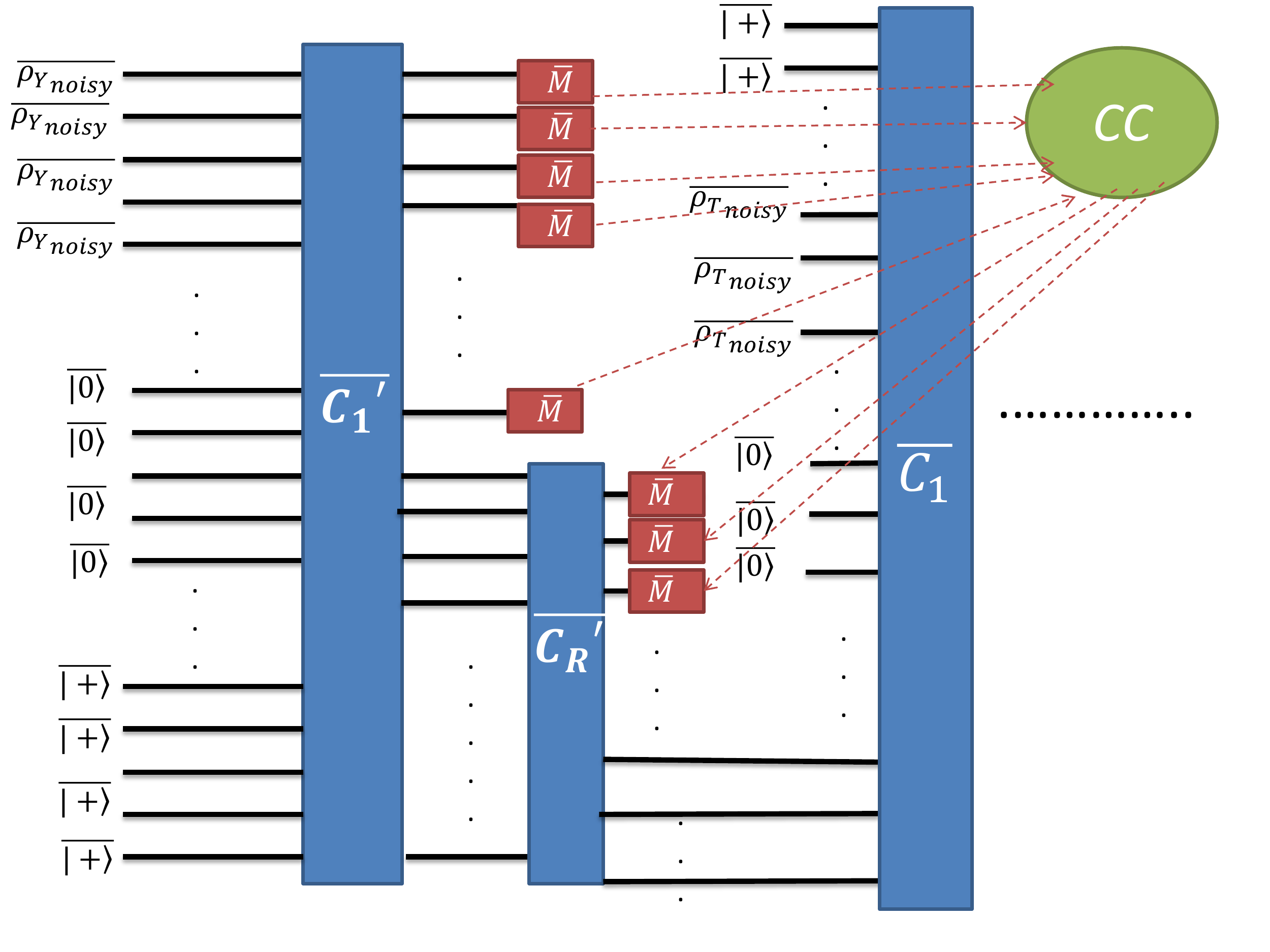}
\caption{Constant depth circuit for our 3D NN architecture. Logical states are up to Pauli corrections due to non-adaptivity. The red box with an $\overline{M}$ symbol is a measurement either in $\overline{X}$ or $\overline{Z}$. Circuit is shown up until $\overline{C_1}$, the remaining part of this circuit is the same as that in Figure \ref{FIG Overview circuit}, with $\overline{Z}$ measurements replaced by $\overline{M}$ measurements, and with some ancilla qubits being initialized in $|\overline{+}\rangle$ as well as $|\overline{0}\rangle$. These slight changes are in order for the construction to be naturally integrated into the RHG framework \cite{RHG07}. Also shown in the figure is the additional interaction with the classical computer $CC$ (ingoing and outgoing red dotted arrows) needed in order to identify the succesfully distilled $Y$-states as well as construct the measurement pattern for the routing circuit $\overline{C^{'}_R}$.}
\label{fig3DNN}
\end{center}
\end{figure*}

 We now very briefly outline the key points in the RHG construction. More detailed explanations can be found in \cite{RHG07,Fujii15,FG08}. In this construction, one starts out with preparing a 3D regular lattice of qubits (call it RHG lattice). This preparation can be done in constant depth by using nearest neighbor $CZ$ gates \cite{RHG07}. This lattice is composed of elementary cells, which can be thought of as smaller 3D lattices building it up. Elementary cells are of two types, primal and dual, and the RHG lattice is composed of a number of interlocked primal and dual cells \cite{RHG07,FG08} . Each elementary cell can be pictured as a cube, with qubits (usually initialized in $|+\rangle$ state) living on the edges and faces of this cube. The RHG lattice is a graph state, and is thus characterized by a set of (local) stabilizer relations \cite{HDE+06}. Errors can be identified by looking at the parity of these stabilizers. Usually, this is done by entangling extra qubits with the systems qubits, these extra qubits are called \emph{syndrome} qubits. However, in the RHG construction this is accounted for by including these syndrome qubits \emph{a priori} when constructing the RHG lattice, this region of syndrome qubits is usually called the \emph{vacuum} region $V$ \cite{RHG07}. Logical qubits in this construction are identified with \emph{defects}. These defects are hole-like regions of the RHG lattice inside of which qubits are measured in the $Z$ basis, effectively eliminating these qubits. Eliminating these qubits (and some of their associated stabilizers) results in extra degrees of freedom which define the logical qubits \cite{RHG07}. Defects can also be primal or dual, depending on whether they are defined on primal or dual lattices. Two defects of the same type (either primal or dual) define a logical qubit. The logical operators $\overline{X}$ and $\overline{Z}$  are products of $X$ operators and $Z$ operators respectively. These products of operators act non-trivially on qubits either encircling each of the two defects, or forming a chain joining the two defects, depending on whether the logical qubit is primal or dual \cite{RHG07,FG08}. By measuring single qubits of the RHG lattice at angles $X$, $Y$, $Z$ and $\dfrac{X+Y}{\sqrt{2}}$, one can perform (primal or dual) logical qubit preparation and measurement in $\overline{X}$ and $\overline{Z}$ bases, preparation of (primal or dual) logical $T$-states and $Y$-states, and logical controlled not ($\overline{CNOT}$) gates between two defects of the same type (this however can only be accomplished by an intermediate step of \emph{braiding} two defects of different types \cite{RHG07}, which is one of the main reasons for the need for two types of defects). If performed perfectly (noiseless case), these operations are universal for quantum computation \cite{RB01}. Note that in our case, as in \cite{KT19}, we will replace measuring qubits in $Y$ and $\dfrac{X+Y}{\sqrt{2}}$ by (equivalently) initializing qubits in $|Y\rangle$ and $|T\rangle$, then measuring these qubits in the $X$ basis. In this way, we will only perform  single qubit $X$ and $Z$ measurements. One of the spatial dimensions of the 3D RHG lattice is chosen as \emph{simulated time}, allowing one to perform a logical version of MBQC via single qubit measurements \cite{RHG07}.

 The preparation and measurement  of logical qubits in the $\overline{X}$ and $\overline{Z}$ bases, as well  $\overline{CNOT}$ can all be performed by measuring qubits of the RHG lattice in $X$ and $Z$ \cite{RHG07,FG08}. All these operations can be performed fault-tolerantly, and non-adaptively (up to Pauli corrections, which  can be pushed until after measurements, and accounted for, since all our circuits are Clifford \cite{RBB03}.), by choosing the defects to have a large enough perimeter, and a large enough separation \cite{FG08,RHG07}. Indeed, in appendix \ref{appRHG1}, we show that when $L_{m}=O(log(n))$, where $L_{m}$ is the minimum (measured in units of length of an elementary cell) between the perimeter of a defect and the separation between two defects in any direction, we would recover the same fault-tolerance results as our 4D NN architecture under local stochastic noise, albeit with different error rates which we will also calculate in appendix \ref{appRHG1}. The noisy logical $Y$-state and $T$-state preparations can also be prepared non-adaptively up to Pauli corrections by performing $X$ and $Z$ measurements on qubits of the RHG lattice, some of which are intialized in $|Y\rangle$ (for logical $Y$-state preparation) or $|T\rangle$ (for logical $T$-state preparation) \cite{FG08}. However, these preparations are unfortunately non-fault-tolerant (introduce logical errors), and therefore these states must be distilled \cite{RHG07}.

 If we could somehow obtain perfect logical $Y$-states, then our constant-depth fault-tolerant 3D NN construction under local stochastic noise would follow a similar analysis as our 4D NN case, and have a  circuit exactly the same as that in Figure \ref{FIG Overview circuit} (up to using $\overline{X}$ measurements in place of $\overline{H}$ gates followed by $\overline{Z}$ measurements), with one difference being that instead of using concatenated versions the of MSD circuits of \cite{HHP+17} to construct $\overline{C_1}$, we will use concatenated versions of the MSD circuits of \cite{HH182}. This is in order to preserve the transversality of logical $T$-gates, which allows preparation of logical $T$-states in the RHG construction by using only local measurements \cite{RHG07} \footnote{This replacement of MSD circuits does not change anything in our proofs, because both of these families of MSD circuits satisfy a specific condition regarding the number of noisy $T$-states needed to distill a $T$-state of arbitrary accuracy, and our proof hinges only on this condition being verified (see appendix \ref{APP subsec zMSD}). Furthermore, the probability  of distillation succeeding for (non-concatenated) MSD circuits in \cite{HH182} approaches one for low enough $\varepsilon$, as in the MSD circuits of \cite{HHP+17}.}. Unfortunately, distilling logical $Y$-states in the RHG construction is essential. What makes matters worse is that using techniques of the likes of those used in the construction of $\overline{C_1}$, on MSD circuits capable distilling logical  $Y$-states up to fidelity $1-\varepsilon_{out}$ (Equation (\ref{eq13PRL})), namely circuits based on the Steane code \cite{RHG07}, leads to circuits with a quasi-polynomial number of ancillas. This is much worse that the polynomial number of ancillas used in circuits $\overline{C_1}$ needed to distill logical $T$-states of the same fidelity $1-\varepsilon_{out}$, and based on the MSD circuits of \cite{HHP+17,HH182} (see appendices \ref{APP subsec zMSD} and \ref{appRHG2}).

 Happily, we manage to overcome this limitation by observing two facts about our construction. The first is that the $Z(\pi/2)$ rotations (and thus $Y$-states) are not needed in order to construct our sampling problem. Indeed, in  Figure \ref{FIG G original} every qubit measured at an $XY$ angle $\pi/2$ in $G_B$ could be replaced by a linear cluster of three qubits measured respectively at  $XY$ angles $\pi/4$, $0$, and $\pi/4$ (these measurements can be implemented  by only  using logical $T$-states in the fault-tolerant version). To make a graph state of regular shape, we should also replace all qubits at the same vertical level as the $\pi/2$-measured qubits in $G_B$ (see Figure \ref{FIG G original}), and which are always measured at an $XY$ angle $0$, with a linear cluster of three qubits measured at an $XY$ angle $0$. By doing this replacement, the new graph gadget $G^{'}_B$ which is an extension of $G_B$ now defines a so-called partially invertible universal set \cite{MGDM19}. Therefore, by results in \cite{MGDM19}, using $G^{'}_B$ instead  of $G_B$ in our construction (Figure \ref{FIG G original}) also results in a sampling problem with distribution $D^{'}=\{D^{'}(s,x)\}$ (where $s$ and $x$ are bit strings defined analogously to those in (\ref{eq3PRL}) and (\ref{eq4PRL})) satisfying both worst-case hardness and the anti-concentration property \cite{MGDM19,HBS+17}. Thus, the distribution $D^{'}$, although different than $D$ ( Equations (\ref{eq3PRL}) and (\ref{eq4PRL})), can be used in the  same way as $D$ to demonstrate a quantum speedup (see Proposition \ref{prop1PRL}). Furthermore, all previous results established for $D$ also hold when $D$ is replaced by $D^{'}$.

 To see why $G^{'}_B$ defines a partially invertible universal set, call $\mathcal{U}_1 \subset U(4)$ ($\mathcal{U}_2 \subset U(4)$) the set of all random unitaries which can be sampled by measuring the qubits of $G_B$ ($G^{'}_B$) non-adaptively at their perscribed angles. Straightforward calculation shows that $ \mathcal{U}_1 \subset \mathcal{U}_2$. Furthermore, both  $ \mathcal{U}_1$ and its complement in $\mathcal{U}_2$ (denoted $\mathcal{U}_2-\mathcal{U}_1$) are (approximately) universal in $U(4)$ since they are composed of unitaries from the gate set of Clifford + T \cite{NC2000,MGDM18}. The set $\mathcal{U}_1$ being both universal in $U(4)$ and inverse containing \cite{MGDM18}, implies that $\mathcal{U}_2$  satisfies all the properties of a partially invertible universal set \cite{MGDM19}. 
 However, note that in using partially invertible universal sets, for technical reasons \cite{MGDM19}, the number of columns of $|G\rangle$ should now satisfy $k=O(n^3)$, resulting in an increase of overhead of ancilla qubits.

One could keep $k=O(n)$ (as in the original construction with $G_B$) while using only $\pi/4$ and $0$ measurements, by using one of the constructions of \cite{BHS+17}. However, the construction of \cite{BHS+17} does not have a provable anti-concentration, although extensive numerical evidence was provided to support the claim that this family of circuits does indeed anti-concentrate \cite{BHS+17}.

 Although $Y$-states are not needed in the construction of our sampling problem, they are still needed to construct MSD circuits for distilling logical $T$-states of fidelity $1-\varepsilon_{out}$ (Equation (\ref{eq13PRL})) \cite{HH182}; which brings us to our second observation. In order to distill logical $T$-states of fidelity $1-\varepsilon_{out}$ (Equation (\ref{eq13PRL})), we only need logical $Y$-states of fidelity $1-\varepsilon^{'}_{out}$ with
 \begin{equation}
     \label{eqepsilonprimee}
     \varepsilon^{'}_{out}=\dfrac{1}{O(poly(log(n)))}.
 \end{equation}
 In other words, the  required output fidelity of the logical $Y$-states need not be as high as that of the logical $T$-states. In appendix \ref{appRHG2}, we show that this leads to a construction of a (constant-depth) non-adaptive MSD (analogous to how  $\overline{C_1}$ is constructed) which takes as input a \emph{polynomial} number of logical ancillas, initialized in either noisy logical $Y$-states, $|\overline{+}\rangle$, or $|\overline{0}\rangle$, and which outputs enough logical $Y$- states of fidelity $1-\varepsilon^{'}_{out}$ needed in the subsequent distillation of logical $T$-states. This circuit, which we call $\overline{C^{'}_1}$ and which is based on concatenations of the Steane code \cite{RHG07}, is a constant depth  Clifford quantum circuit composed of $\overline{CNOT}$ gates, and followed by non-adaptive $\overline{X}$ and $\overline{Z}$ measurements. $\overline{C^{'}_1}$, as $\overline{C_1}$, prepares the graph states needed for non-adaptive MSD via MBQC (as seen previously). Note that here we will use $\overline{CNOT}$ gates instead of $\overline{CZ}$ gates in order to prepare logical graph states, since these gates are more natural in the RHG construction \cite{RHG07}. The preparation procedure is essentially the same as that with $\overline{CZ}$ modulo some $\overline{H}$ gates, but these logical Hadamards can be absorbed into the initialization procedure (where some qubits become initialized in $|\overline{0}\rangle$ instead of $|\overline{+}\rangle$) and the measurements (where some $\overline{X}$ measurements after $\overline{C^{'}_1}$ are changed to $\overline{Z}$ measurements, and vise versa.). The same holds for all other circuits based on graph states in this construction.

 With the distillation of logical $Y$-states taken care of, we now summarize our constant depth construction based on a 3D NN architecture. The circuit of this construction is found in Figure \ref{fig3DNN}. It takes as input logical qubits initialized in the states $|\overline{+}\rangle$, $|\overline{0}\rangle$, $\overline{\rho_T}_{noisy}$, and $\overline{\rho_Y}_{noisy}$, and outputs a bit string $(\overline{s},\overline{x})$ sampled from the distribution $\tilde{\overline{D^{'}}}_2$ demonstrating a quantum speedup  (see Theorem \ref{TH1} and Proposition \ref{prop1PRL}). Note that $\tilde{\overline{D^{'}}}_2$ is the fault-tolerant version of the distribution $D^{'}$ defined earlier. $\tilde{\overline{D^{'}}}_2$ is defined analogously to $\tilde{\overline{D}}_2$ in Equation (\ref{eq14PRL}), which is the fault-tolerant version of the distribution $D$ (Equation (\ref{eq4PRL})). Our 3D NN architecture is composed of five constant depth circuits acting on logical qubits, $\overline{C^{'}_1}$, $\overline{C^{'}_R}$, $\overline{C_1}$, $\overline{C_R}$, and $\overline{C_2}$. $\overline{C^{'}_1}$, $\overline{C_1}$, $\overline{C_R}$, and $\overline{C_2}$ are as defined previously, and  $\overline{C^{'}_R}$ is a routing circuit, analogous to $\overline{C_R}$, which routes succesfully distilled logical $Y$-states to be used in $\overline{C_1}$. Furthermore, all of these circuits, as well as the preparation of logical qubits, can be constructed by non-adaptive single-qubit $X$ and $Z$ measurements on physical qubits arranged in a 3D RHG lattice, whose preparation is constant depth and involves only nearest neighbor $CZ$  gates. These physical qubits are initialized in the (noisy) states $|+\rangle$, $|Y\rangle$, and $|T\rangle$ \cite{RHG07}. Our construction has two layers of interaction with a classical computer, needed to identify succesfully distilled logical $Y$ and $T$-states respectively. The number of physical qubits needed is $O(n^{11}poly(log(n))$, this calculation is performed in appendix \ref{appRHG2}. The additional  overhead as compared to our 4D NN construction comes from mainly two sources, the partially invertible universal set condition \cite{MGDM19}, and the circuits $\overline{C^{'}_R}$ and $\overline{C^{'}_1}$ which arise as a result of needing to distill logical $Y$-states in the 3D RHG construction \cite{RHG07}.

 As in our 4D NN architecture, the noise model we use here is  the local stochastic quantum noise defined earlier \cite{BGK+19,fawzi2018constant}. Since the circuit needed to construct the 3D RHG lattice is composed of single and two-qubit Clifford gates acting on prepared qubits \cite{RHG07}, all errors of preparations and gate applications can be pushed, together with the measurement errors, until after the measurements; as seen previously. Because the circuit preparing the RHG lattice is constant depth, the overall local stochastic noise has a constant rate (see Equation (\ref{eq7PRL})), and therefore could be corrected with high probability for low enough (constant) error rates of preparation, gate application, and measurements \cite{BGK+19} (see appendix \ref{appRHG1} where we calculate an estimate of these error rates). The error correction, as in our 4D NN architecture, is completely classical and involves minimal weight matching \cite{edmonds1973operation}. This error correction is $poly(n)$-time and is performed at each of the two layers of interaction with the classical computer, as well as after the final measurements. Also, as in the 4D NN case, other $poly(n)$-time classical algorithms are included in the classical post processing; these are in order to identify succesful MSD instances, and identify the measurement patterns of the routing circuits. The classical computer at each layer of interaction as well as after the final measurements takes as input measurement results of qubits involved in the computation, as well as measured qubits in the vacuum region $V$. These vacuum qubits give the error syndrome at multiple steps in the computation, and are therefore needed for the minimal weight matching \cite{RHG07}.

\bigskip
\bigskip

$\textbf{\emph{Discussion}}-$ In summary, we have presented a construction sampling from a distribution  demonstrating a quantum speedup, which is robust to noise. 
Our construction has constant depth in its quantum circuit, and can be thought of as a fault-tolerant version of the (noise free) constant depth quantum speedup based on generating and measuring graph states \cite{gao17,BHS+17,HBS+17,MGDM18,MGDM19,HHB+19}. We have shown how to implement this construction both by using a 4D architecture with nearest neighbor two-qubit gates, or by using a 3D architecture with nearest neighbor two-qubit gates. The circuits of each of these architectures interact at most twice with an (efficient) classical device while running, and  have different requirements in terms of overhead of physical ancilla qubits, owing to the fact that they are based on two different constructions for fault-tolerance \cite{BGK+19,RHG07}. 


The overheads are large in terms of the number of (physical) qubits, however these may be improved. In any case, our construction is considerably simpler than fault-tolerant full blown quantum computation where circuits are scaling in depth and many adaptive layers are required. Therefore our architectures demonstrate potential for interim demonstration of quantum computational advantage, which may be much more practical. 
Indeed, if one considers classical computation temporally free, our construction represents a constant time implementation of a sampling problem with fault-tolerance.


We note that although we have presented here a fault-tolerant construction for a specific graph state architecture \cite{MGDM18}, the same techniques can be applied to any of the sampling schemes based on making local $XY$ measurements from the set $\{0,\pi/2,\pi/4\}$  on  regular graph states \cite{gao17,BHS+17,HBS+17,MGDM18,MGDM19}.

In particular it can be easily adapted to cases where the measurements are not fixed but chosen at random before the running of the circuit \cite{BHS+17,HBS+17,gao17}. This would essentially just fix the locations of the distilled $T$-states, but it could be done before hand, and would not effect the efficiency of the routing circuits. This has the potential of relating the average-case hardness conjecture to that of other more familiar problems \cite{gao17,BHS+17,BMS16PRL}.

Our work also has potentially  another interest, as it can alternatively be viewed as a constant depth quantum circuit which samples from an approximate unitary $t$-design \cite{DCE+09} fault-tolerantly. Indeed, our techniques can be used to directly implement a logical version of Equation (\ref{eq2PRL}), which samples from an approximate $t$-design. These $t$-designs have many useful applications across quantum information theory \cite{DCE+09,emerson2003pseudo,hayden2004randomizing,matthews2015testing,muller2015thermalization,hayden2007black}.


Several interesting approaches for optimization may be considered. One could think of using different quantum error correcting codes, such as those of \cite{LAR+11,fawzi2018constant}, to decrease the overhead of physical qubits. One could also aim to optimize the overhead of both gates and physical qubits of the MSD by using techniques similar to those of \cite{CC19,CN20}.

The ability to efficiently verify quantum speedup is also an important goal. Although this question has already been pursued in the regime of fault-tolerance in \cite{KT19}, and the techniques developped there are directly applicable to our 3D NN architecture; it would be interesting to develop verification techniques more naturally tailored to the graph state approach \cite{HDE+06,RB01} and MBQC \cite{RB01,RBB03}, which we use heavily here.  In this direction, the work of \cite{MK18,TMM+19} can  be used for this purpose when the measurements (both Clifford and non-Clifford) as well as the $CZ$ and Hadamard gates (needed for the preparation of the graph states \cite{HDE+06}) are assumed $perfect$ (noiseless). Indeed, in this case the verification amounts to verifying that the graph state was correctly prepared, for which \cite{MK18,TMM+19} provide a natural path to do so, by giving good lower bounds (with high confidence) on the fidelity (with respect to the ideal graph state corresponding to the sampling problem) of the prepared graph state in the case where a sufficient amount of stabilizer tests pass \cite{MK18,TMM+19}. These lower bounds on the fidelity, tending asymptotically to one \cite{MK18,TMM+19}, allow one to verify that quantum speedup is being observed, as long as one trusts the local measurement devices (which, being small, can be checked by other means efficiently). This verification of quantum speedup can be done by using the standard relation between  the fidelities of two quantum states (which in our case are the ideal state and the state accepted by the verification protocol) and the $l_1$-norm of the two output probability distributions corresponding to measuring the qubits of these two states \cite{NC2000}.

These techniques, however, do not easily extend to the case where the measurements and gates needed for preparation are noisy; since for graph states of size $m$, even for an arbitrarily small (but $constant$, for example below the threshold for fault-tolerant computing) noise strength, the verification protocol might fail (not accept a good state) in the asymptotic ($m \to \infty$) limit (see for example \cite{TMM+19} where the verification accepts with probability one asymptotically only if the noise strenght scales as $1/poly(m)$). We leave this problem  for future investigation.


 


 
\bigskip

\begin{acknowledgments}
We thank David Gosset, Elham Kashefi, Theodoros Kapourniotis, Anthony Leverrier, Ashley Montanaro, Micha\l{} Oszmaniec, and Peter Turner for fruitful discussions and comments. The Authors would like to acknowledge the National Council for Scientific Research
of Lebanon (CNRS-L) and the Lebanese University (LU) for granting a doctoral fellowship to R. Mezher.
We acknowledge support of  the ANR
through the ANR-17-CE24-0035 VanQute project.
\end{acknowledgments}
\nocite{*}
\bibliography{apssamp}

\onecolumngrid

\appendix

\section{Size of encoding and intermediate case hardness of sampling.}
\label{APPC}

Here we prove our statements regarding the sufficiency of the size of logical encoding $l$ (Equation (\ref{eq10PRL})) and the proof of hardness in the intermediate case where we have noise in the circuit, but assume perfect $T$-states (Equation (\ref{eq11PRL})).
As mentioned in the main text, the probability $p_f$ that the classical decoding fails to correct an error $E \sim \mathcal{N}(p)$ affecting a surface code composed of $l$ physical qubits is given by \cite{DKL+02,BGK+19,fowler2012proof}
\begin{equation}
    \label{eqappC1}
    p_f=e^{-O(cd)}=e^{-O(\sqrt{l})},
\end{equation}
when the error rate $p$ is below the threshold for fault-tolerant computing with the surface code \cite{DKL+02}. We will assume, as mentioned in the main text, that the error rates of preparation, single and two qubit gates, and measurements in our construction are small enough, that is, below the threshold of fault-tolerant computing with the surface code, and classical postprocessing is instantaneous. We will also assume that the probabilities of failure of the classical decoding algorithms in each logical qubit are independent (see \footnote{This assumption may seem strong, but it is actually very mild and has no effect on our end result. To see this, suppose we drop this assumption, then the probability of success of all $k.n$ decodings should now be calcuated by a union bound. From the properties of local stochastic noise (namely that local stochastic noise on a subset of qubits of the system is still local stochastic with the same rate \cite{BGK+19}) a decoding of a logical qubit succeeds (is able to identify and correct for the error) with probability $p_{single}=1-p_f=1-e^{-O(\sqrt{l})}$ (when the error rates of all local stochastic noise in our construction are adequately low, i.e below the threshold of fault-tolerant computing with the surface code), therefore the probability that all $k.n$ decodings succeed is given by $P=1-k.n+k.n.p_{single}=1-k.n.e^{-O(\sqrt{l})}$, by a standard bound on the intersection of $k.n$ events derived from a union bound. The assumption we make in the main text results in a good approximation of $P$, and is simpler to state (which is why we used it in the main text). Finally, note that this does not mean that errors between physical qubits of two entangled logical qubits are uncorrelated. Indeed, the correlation between  these qubits is accounted for in the propagation rules of local stochastic noise \cite{BGK+19}, since forward propagating local stochastic noise  in  Clifford circuits composed of single and two-qubit gates generally results in local stochastic noise with higher error rate \cite{BGK+19}.}). Our construction involves classical decoding of the measurement results of $O(k.n)$ logical qubits \footnote{Actually, it is something like $2.k.n$ if we include decoding of measured logical qubits of the Bell states obtained at the end of the single shot procedure of \cite{BGK+19} (see $[57]$). This changes nothing in the analysis we have done, so we chose to omit it in the main text for simplicity.}. After decoding, the probability of observing outcome $(\overline{s},\overline{x})$ is given by
\begin{equation}
    \label{eqappC2}
    \tilde{\overline{D}_1}(\overline{s},\overline{x})=(1-e^{-O(\sqrt{l})})^{O(k.n)}\overline{D}(\overline{s},\overline{x})+\sum_{i}p_{i}\overline{D}_i^{e}(\overline{s},\overline{x})
\end{equation}
where $\overline{D}_i^{e}$ is a distribution corresponding to sampling from the outputs $(\overline{s},\overline{x})$ of $|\overline{G'}\rangle$ in the presence of local stochastic noise, and where the decoding algorithm has failed in at least one logical qubit. $\sum_{i}p_i\overline{D}_i^{e}(\overline{s},\overline{x})$ enumerates all possible ways in which decoding on  the $k.n$ logical qubits of $|\overline{G'}\rangle$ can fail. Note that 
\begin{equation*}
    \sum_{i}p_i= 1-(1-e^{-O(\sqrt{l})})^{O(k.n)}.
\end{equation*}
Now,
\begin{equation}
\label{eqappCf}
    \sum_{\overline{s},\overline{x}}|\tilde{\overline{D}_1}(\overline{s},\overline{x})-\overline{D}(\overline{s},\overline{x})| = \sum_{\overline{s},\overline{x}}|(1-e^{-O(\sqrt{l})})^{O(k.n)}\overline{D}(\overline{s},\overline{x})+\sum_{i}p_{i}\overline{D}_i^{e}(\overline{s},\overline{x})-\overline{D}(\overline{s},\overline{x})| \leq 2(1-(1-e^{-O(\sqrt{l})})^{O(k.n)}).
\end{equation}
The bound on the right hand side is obtained from a triangle inequality and by noting that $\sum_{\overline{s},\overline{x}}\overline{D}(\overline{s},\overline{x})=\sum_{\overline{s},\overline{x}}\overline{D}_i^{e}(\overline{s},\overline{x})=1$.
Choosing
\begin{equation}
    \label{eqblocksize}
    l=r.log^{2}(n)=O(log^2(n)),
\end{equation}
where $r$ is a positive constant chosen large enough so that the following inequality holds
\begin{equation}
    \label{eqcondblocksize}
    deg(e^{O(\sqrt{l})}) > deg(k.n),
\end{equation}
where $deg(.)$ represents the highest power of $n$  in the expressions of $e^{O(\sqrt{l})}$ and \\ $O(k.n)$. We can now use (for large enough $n$) the approximation 
\begin{equation}
\label{eqappCapprox}
2\big(1-(1-e^{-O(\sqrt{l})})^{k.n})  \sim 2e^{-O(\sqrt{l})}.k.n=O(\dfrac{1}{n^{\beta}}),
\end{equation}
with $\beta=deg(e^{O(\sqrt{l})})-deg(k.n)$. Plugging Equation (\ref{eqappCapprox}) in Equation (\ref{eqappCf}) we get
\begin{equation*}
\sum_{\overline{s},\overline{x}}|\tilde{\overline{D}_1}(\overline{s},\overline{x})-\overline{D}(\overline{s},\overline{x})| \leq O(\dfrac{1}{n^{\beta}})=\dfrac{1}{poly(n)}.
\end{equation*}
This completes the proof of Equations (\ref{eq10PRL}) and (\ref{eq11PRL}).

\section{Bounding $\tilde{\overline{D}}_2(\overline{s},\overline{x})$ and Properties of $zMSD$}
\label{app zMSD}

\subsection{Bounding $\tilde{\overline{D}}_2(\overline{s},\overline{x})$ (proof of Equation (\ref{eq14PRL}))}
Let
\begin{equation}
    \label{eqappD1}
    \tilde{\rho}_{|\overline{G'}\rangle}=\bigotimes_{a\in V} \overline{H}_a \bigotimes_{b \in V_2}\overline{Z(\pi/2)}_b \prod_{\{i,j\}\in E}\overline{CZ}_{ij} \bigotimes_{c\in V/V_1}\overline{H}_c\overline{|0}\rangle_c \langle \overline{0}|_c\overline{H}_c \bigotimes_{d\in V_1}\overline{\rho_T}_{out}  \prod_{\{i,j\}\in E}\overline{CZ}_{ij}\bigotimes_{b \in V_2}\overline{Z(-\pi/2)}_b \bigotimes_{a\in V} \overline{H}^{\dagger}_a.
\end{equation}
$\tilde{\rho}_{|\overline{G'}\rangle}$ is exactly the same as $|\overline{G'}\rangle$, but with each single logical qubit state $|\overline{T}\rangle$ replaced with $\overline{\rho_T}_{out}$, the output of a succesful instance of $zMSD$ (Equations (\ref{eqPRL12}) and (\ref{eq13PRL})).
The probability $\tilde{\overline{D}}_2(\overline{s},\overline{x})$ can be calculated by using the following simple observation
\begin{equation}
 \label{eqappD2}
 \tilde{\overline{D}}_2(\overline{s},\overline{x})=p(\{\overline{s},\overline{x}\}\cap ne)+p(\{\overline{s},\overline{x}\}\cap e),
\end{equation}
where $p(\{\overline{s},\overline{x}\}\cap ne)$ is the probability of observing outcome $\{\overline{s},\overline{x}\}$  when no logical error (ne) has occured (that is, that  classical decoding did not fail in any logical qubit) , neither in the distillation process, nor in the routing, nor in constructing and measuring $\tilde{\rho}_{|\overline{G'}\rangle}$. $p(\{\overline{s},\overline{x}\}\cap e)$ is the probability of observing $\{\overline{s},\overline{x}\}$ when the  decoding algorithm has failed (e) at least on one logical qubit. We will assume that in the  case where no logical error has occured, for large enough $n$, the probability $p_{succ}$ (Equation (\ref{eqpsuccprl})) of distilling enough ($O(n^2))$ states $\overline{\rho_T}_{out}$  to construct $\tilde{\rho}_{|\overline{G'}\rangle}$ is equal to one. This is a reasonable assumption since the exponential term in $p_{succ}$
varies much more rapidly than the polynomial terms in our bounds, for large enough $n$. Now,
\begin{equation*}
  p(\{\overline{s},\overline{x}\}\cap ne)=p(ne).p( \{\overline{s},\overline{x}\}| ne).
  \end{equation*}
  \begin{equation}
  \label{eqpne}
      p(ne)=(1-e^{-O(\sqrt{l})})^{O(n^5log^2(n))},
  \end{equation}
is the probability that that the decoding does not fail on all our $O(n^5log^2(n))$ logical qubits (logical qubits of all copies of $zMSD$, the routing circuit, as well as 
$\tilde{\rho}_{|\overline{G'}\rangle}$). 
Now,
\begin{equation}
\label{eqpsxne}
    p(\{\overline{s},\overline{x}\}|ne)=\sum_{i_1,...,i_{k.n.l}}\langle i_1...i_{k.n.l} |\tilde{\rho}_{|\overline{G'}\rangle}| i_1...i_{k.n.l}\rangle,
\end{equation}
where $|i_1...i_{k.n.l}\rangle$ is a state  of $k.l.n$ physical qubits, corresponding to the measurement of the $k.n$ logical qubits of $\tilde{\rho}_{|\overline{G'}\rangle}$, which when decoded gives rise to the bit string $(\overline{s},\overline{x})$.   
\begin{equation}
\label{eqpsxe}
    p(\{\overline{s},\overline{x}\}\cap e)=\sum_{j}p_{e_j}p(\{\overline{s},\overline{x}\}|e_j),
\end{equation}
where the right hand of Equation (\ref{eqpsxe}) enumerates all possible ways in which decoding on the $O(n^5log^2(n))$ logical qubits could fail.
Note that 
\begin{equation}
    \label{eqboundpsxe}
    \sum_{\overline{s},\overline{x}}p(\{\overline{s},\overline{x}\}\cap e) \leq \sum_jp_{e_j} \leq 1-p(ne) \leq 1-(1-e^{-O(\sqrt{l})})^{O(n^5log^2(n))}.
\end{equation}
Replacing Equations (\ref{eqpne})-(\ref{eqpsxe}) in Equation (\ref{eqappD2}) we get

\begin{equation}
    \label{eqappD7}
    \tilde{\overline{D}}_2(\overline{s},\overline{x})=(1-e^{-O(\sqrt{l})})^{O(n^5log^2(n)))}p(\{\overline{s},\overline{x}\}|ne)+\sum_{j}p_{e_j}p(\{\overline{s},\overline{x}\}|e_j).
\end{equation}
By using Equations (\ref{eqboundpsxe}) and (\ref{eqappD7}) as well as a triangle inequality. We get that
\begin{equation}
    \label{eqappD8}
    \sum_{\overline{s},\overline{x}}|\tilde{\overline{D}}_2(\overline{s},\overline{x})-p(\{\overline{s},\overline{x}\}|ne)| \leq 2(1-(1-e^{-O(\sqrt{l})})^{O(n^5log^2(n))}).
\end{equation}
As in appendix \ref{APPC}, chosing $$l =r.log^2(n),$$
but  with $r$ chosen so that $$deg(e^{O(\sqrt{l})}) > deg(O(n^5log^2(n))),$$
we get that 
\begin{equation}
\label{eqappD9}
  \sum_{\overline{s},\overline{x}}|\tilde{\overline{D}}_2(\overline{s},\overline{x})-p(\{\overline{s},\overline{x}\}|ne)| \leq \dfrac{1}{poly(n)}  
\end{equation}
by using the same approximations as in appendix \ref{APPC} to bound $2(1-(1-e^{-O(\sqrt{l})})^{O(n^5log^2(n))})$.
Now, remark that the fidelity between $\tilde{\rho}_{|\overline{G'}\rangle}$ and $|\overline{G'}\rangle$, denoted as $F$, satisfies (from Equations (\ref{eqappD1}) and (\ref{EQN: logical G'})) 
\begin{equation}
\label{eqappD10}
    F \geq (1-\varepsilon_{out})^{O(n^2)},
\end{equation}
with $\varepsilon_{out}$ given by Equation (\ref{eq13PRL}). Furthermore, the probabilities $\overline{D}(\overline{s},\overline{x})$
and $p(\{\overline{s},\overline{x}\}|ne)$
satisfy \cite{NC2000} \footnote{Actually, this relation holds for the $l_1$-norm distance between the probability distributions over physical qubits. However, as the absolute value of the sum is less than the sum of absolute values, this relation also holds for the probabilities in Equation (\ref{eqappD11}).}
\begin{equation}
\label{eqappD11}
    \sum_{\overline{s},\overline{x}}|\overline{D}(\overline{s},\overline{x})-p(\{\overline{s},\overline{x}\}|ne)| \leq 2\sqrt{1-F^2}.
\end{equation}
when $\varepsilon_{out}$ satisfies Equation (\ref{eq13PRL}), 
$$2\sqrt{1-F^2} \leq  2\sqrt{1-(1-\varepsilon_{out})^{O(n^2)}} \sim 2 \sqrt{O(n^2)\varepsilon_{out}} \leq \dfrac{1}{poly(n)}.$$
Plugging this into Equation (\ref{eqappD11}), then using Equations (\ref{eqappD9}) and (\ref{eqappD11}) and a triangle inequality, we obtain
\begin{equation*}
    \sum_{\overline{s},\overline{x}}|\tilde{\overline{D}_2}(\overline{s},\overline{x})-\overline{D}(\overline{s},\overline{x})| \leq \dfrac{1}{poly(n)}.
\end{equation*}
This completes the proof of Equation (\ref{eq14PRL}).

\subsection{Properties of $zMSD$} \label{APP subsec zMSD}
$zMSD$ implements non-adaptively $z$ iterations of the MSD protocol of Theorem 4.1 in \cite{HHP+17}. Note that in the protocol of \cite{HHP+17}, the MSD circuit was for magic states of the form $|H\rangle=cos(\pi/8)|0\rangle+sin(\pi/8)|1\rangle$ whereas in our case we need distillation circuits for $T$-states $|T\rangle $ defined in the main text. However, since $HZ(-\pi/2)|H\rangle=e^{-i\pi/8}|T\rangle$, the circuits in \cite{HHP+17} can be adapted to our case by adding a constant depth layer of $H$ and $Z(-\pi/2)$ gates, whose logical versions can be done fault-tolerantly and also in constant depth in our construction . We call $1MSD$ a circuit which implements non-adaptively one iteration of the protocol of Theorem 4.1 in \cite{HHP+17}. Note that both $zMSD$ and $1MSD$ will be based on non-adaptive  MBQC. We will begin by calculating the number of qubits of $1MSD$.

\begin{figure}[h]
\begin{center}
\graphicspath{}
\includegraphics[trim={2 0cm 150 0cm} , scale=0.5]{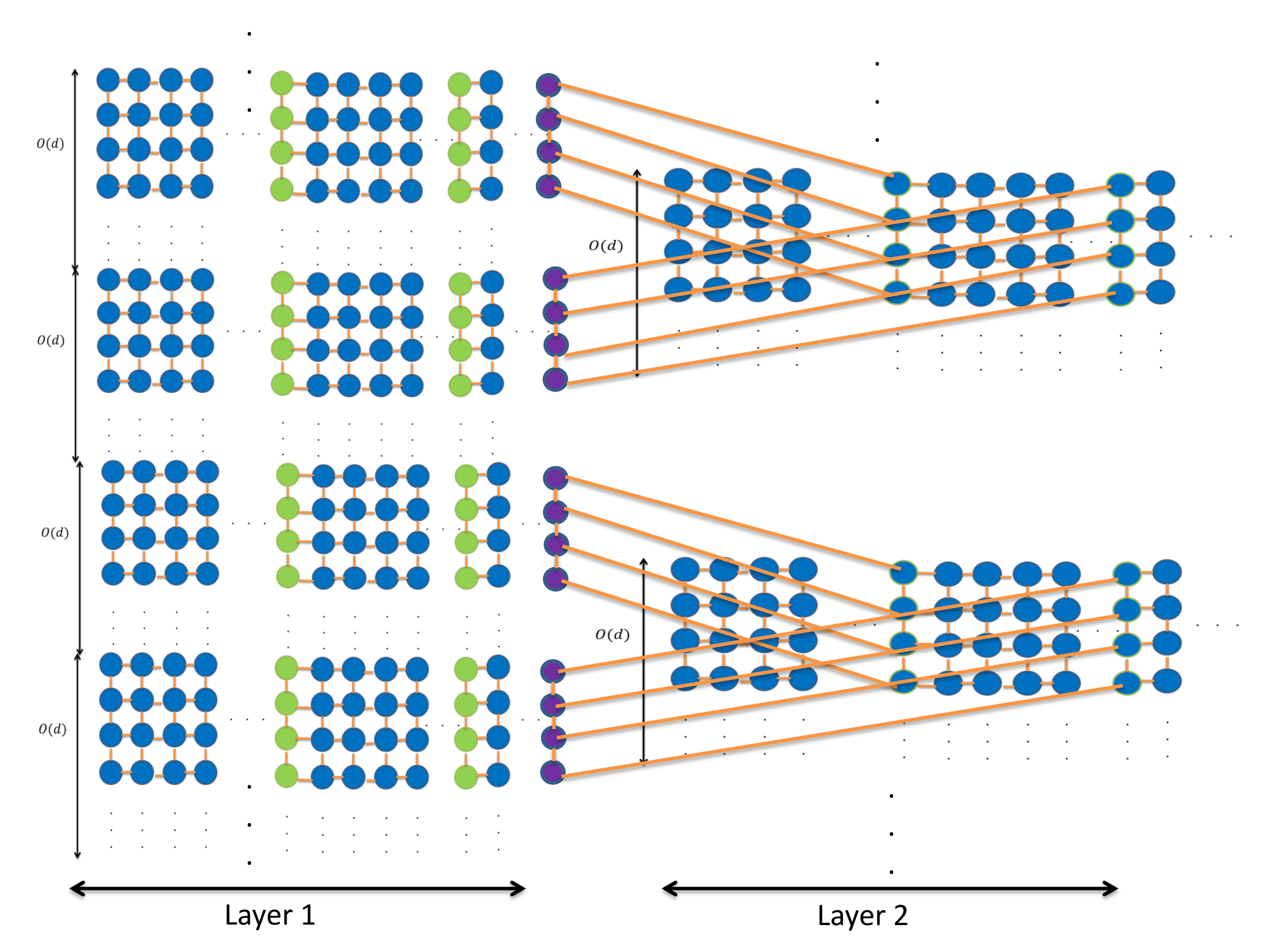}

\caption{Part of the graph state implementing the circuit $zMSD$. Blue filled circles represent logical qubits in the $|\overline{+}\rangle$ state, which when measured implement the Clifford part of the MSD protocol of Theorem 4.1 in \cite{HHP+17}. The green filled circles are noisy input $T$-states $\overline{\rho_T}_{noisy}$. Purple filled circles are the output qubits of the first layer of $zMSD$. When  $zMSD$ is successful, these qubits are in a state with fidelity $1-O(\varepsilon^d)$ with respect to the ideal $T$-state $|\overline{T}\rangle$. The orange lines are $\overline{CZ}$ gates. Note that the output qubits of the first layer (purple circles) are connected to the second layer at different positions by means of long range $\overline{CZ}$ gates. These long range $\overline{CZ}$ gates can be implemented in constant depth, since they act each on distinct pairs of qubits. Also, as mentioned in the main text in this appendix, these long range $\overline{CZ}$ gates can be replaced by a series of  $\overline{SWAP}$ gates making this construction a constant-depth 2D construction with only nearest-neighbor $\overline{CZ}$ gates. Measurements consist of non-adaptive $\overline{X}$ measurements, $\overline{Z}$ measurements, as well as $\overline{Y}$ measurements. As described in the main text, we could equivalently perform all measurements in $\overline{Z}$, by introducing additional constant depth layers of $\overline{H}$ and $\overline{Z(\pi/2)}$ gates.}
\label{fig1}
\end{center}
\end{figure}

In Theorem 4.1 in \cite{HHP+17}, the MSD circuit takes as input $O(d)$ qubits, where $d$ is a positive integer, uses $O(d^2)$ noisy input $T$-states with fidelity 1-$\varepsilon$ with respect to an ideal (noiseless) $T$-state, and outputs $O(d)$ distilled $T$-states with fidelity 1-$O(\varepsilon^{d})$ with respect to an ideal $T$-state (note that the ratio of the number of noisy input $T$-states to the number of distilled output $T$-states is $\sim d$ for large enough constant $d$ \cite{HHP+17}.). Each time a noisy $T$-state is inserted it affects a noisy $T$-gate, inducing a so-called $T$-gate depth \cite{HHP+17}. The depth of the entire circuit is $O(d^2.log(d))$, where $O(d)$ is the $T$-gate depth, and $O(d.log(d))$ is the depth of the Clifford part of the circuit, which is composed of long-range Cliffords \cite{HHP+17}. Therefore, the MSD circuit is an $O(d)$-qubit circuit of depth $O(d^2.log(d))$. In order to implement this circuit on a regular graph state (for example, the cluster state \cite{RB01}), one must transform the Clifford circuit composed of long range gates, to that composed of nearest neighbor and single qubit Clifford gates, since these single qubit and nearest neighbor two-qubit gates can  be implemented by measuring $O(1)$  qubits of a cluster state in the $X$ and $Y$ bases  \cite{RB01,RBB03}. An $m$-qubit Clifford gate can be implemented by an $O(m^2)$-depth circuit composed only of  gates from the set  $\{CZ_{ij},H,Z(\pi/2)\}$ \cite{gottesman1997stabilizer}. Furthermore, $CZ_{ij}$ could be implemented by a circuit of depth $O(i-j)$ composed of nearest neighbor CZ gates \cite{mantri2017universality}. The same arguments hold in the logical picture by replacing $H$, $CZ$, $Z(\pi/2)$, and noisy input $T$-states with their logical versions $\overline{H}$, $\overline{CZ}$,  $\overline{Z(\pi/2)}$, and $\overline{\rho_T}_{noisy}$.
$m=O(d)$ in our case, thus the number of columns of the cluster state needed to implement $1MSD$ is
\begin{equation}
    \label{eqnumberofcolumns}
    n_{c}=O(d^2log(d)).O(d^2).O(d)=O(d^5log(d)),
\end{equation}
where the $O(d^2log(d))$ comes from the depth of the MSD circuit with long range Cliffords, $O(d^2)$ is the depth needed to implement an arbitrary  Clifford using $\overline{H}$ gates, $\overline{Z(\pi/2)}$ gates, and long range $\overline{CZ}$'s, and the $O(d)$ is an overestimate and represents the  number of nearest neighbor $\overline{CZ}$'s needed to give a long range $\overline{CZ}$. The total number of qubits of the cluster state implementing $1MSD$ is then
\begin{equation}
    \label{eqtotalqubits}
    n_{T}=O(d).n_{c}=O(d^{6}.log(d)).
\end{equation}
$zMSD$ can be thought of as a concatenation of $z$ layers of $1MSD$, where the output of layer $j$ is the input of layer $j+1$. Because the noisy input $T$-states in the protocol of \cite{HHP+17} are injected at different parts of the circuit, this means that the output qubits of layer $j$ should be connected to layer $j+1$ at different positions by means of long range $\overline{CZ}$ gates. Therefore, the graph state implementing $zMSD$ can be seen as cluster states composed of logical qubits, and connected by long range $\overline{CZ}$ gates, as shown in Figure (\ref{fig1}).  One could equivalently replace these long range $\overline{CZ}$ gates with a series of $\overline{SWAP}$ gates, which can be implemented (up to Pauli correction by means of non-adaptive $\overline{X}$ and $\overline{Z}$ measurements) on a 2D cluster state with only nearest neighbor $\overline{CZ}$ gates \cite{RB01,RBB03}. Because these long range $\overline{CZ}$ gates act on qubits separated by a distance $poly(d)$, the introduction of $\overline{SWAP}$ gates introduces an additional (constant) overhead of $O(poly(d))$ qubits to  $n_T$, but makes the construction of $1MSD$ implementable on a 2D cluster state with only nearest neighbor $CZ$ gates.
The first layer consists of $N$ copies of cluster states implementing $1MSD$ (see Figure \ref{fig1}), and outputs, when succesful, $N.O(d)=\dfrac{N}{d}.O(d^2)$ $T$-states with fidelity $1-C.\varepsilon^d$ with respect to $|\overline{T}\rangle$, $C$ being a positive constant \cite{HHP+17}. These $T$-states are the input of the second layer which consists of $\dfrac{N}{d}$ copies of cluster states implementing $1MSD$, and outputs, when succesful, $\dfrac{N}{d}.O(d)=\dfrac{N}{d^2}.O(d^2)$ $T$-states with fidelity $C.(C.\varepsilon^{d})^{d}=C^{d+1}.\varepsilon^{d^2}$ with respect to $|\overline{T}\rangle$. Similarly, 
the $z$th layer will consist of $\dfrac{N}{d^{z-1}}$ copies, and will output, when successful,  $\dfrac{N}{d^{z-1}}.O(d)$ $T$-states with fidelity 
\begin{equation}
\label{eqepsilonout}
\varepsilon_{out} \sim C^{d^{z-1}}.\varepsilon^{d^z},
\end{equation}
with respect to $|\overline{T}\rangle$.
The total number of qubits of the graph state implementing $zMSD$ is then given by
\begin{equation}
    \label{eqtotalqubitsnmsdl}
    n_{NMSD}=(N+\dfrac{N}{d}+\dfrac{N}{d^2}+...).n_T=O(N).
\end{equation}
$z$ is the last layer, therefore $\dfrac{N}{d^{z-1}} = 1$ and thus
\begin{equation}
\label{eqN1}
    N = d^{z-1}.
\end{equation}
For a succesful instance of $zMSD$, in order to arrive at Equation (\ref{eq13PRL}), choose $$d^z \geq O(log(n)),$$ this implies that each copy of $zMSD$ is composed of 
$$n_{NMSD}=O(N)=O(d^{z-1}) \geq O(log(n)),$$
logical qubits, as mentioned in the main text. Indeed, replacing  $d^z = a.log(n),$ with $a$ a positive constant in Equation (\ref{eqepsilonout}) yields 
$$\varepsilon_{out} = \dfrac{1}{n^{a.\alpha}},$$
by a direct calculation, where $\alpha=\dfrac{log(\dfrac{1}{C.\varepsilon^d})}{d}$ while noting that $C.\varepsilon^d < 1$ \cite{HHP+17}. Equation (\ref{eq13PRL}) is therefore obtained for an appropriate choice of $a$ or $\varepsilon$. 

Now, we will calculate the probability $p_{szMSD}$ of  a single successful instance of $zMSD$. We will assume, rather pessimistically, that only one string of non-adaptive measurement results of $zMSD$ corresponds to a successful instance. This string we will take, by convention, to be the one where all the  measurement binaries (after decoding ) are zero. In this case,
\begin{equation}
    \label{eqpszMSD}
    p_{szMSD} \geq \dfrac{1}{2^{n_{NMSD}}}.
\end{equation}
Note that the lower bound is actually higher than that Equation (\ref{eqpszMSD}) for two reasons. The first is that not all qubits of the graph state implementing $zMSD$ are measured. Indeed, the output qubits of the last layer of $zMSD$ are unmeasured and, in the case when $zMSD$ is successful, are in the state $\overline{\rho_T}_{out}$. The second reason is that some of the measurements correspond, in the successful case, to post-selections which in the protocol of \cite{HHP+17} occur with probability greater  than $1/2$ .  Indeed, for small enough $\varepsilon$ , the acceptance rate of the protocol of \cite{HHP+17}   is approximately 1.
Now, $\varepsilon_{out} = \dfrac{1}{n^{\beta}}$, with $\beta \geq 4$, and $n_{NMSD}=\gamma.d^z$ (Equations (\ref{eqtotalqubitsnmsdl}) and (\ref{eqN1})), with $\gamma$ a positive constant. By choosing  $$\varepsilon=\dfrac{e^{-\gamma.\beta.log(2)}}{C^{1/d}},$$ and performing a direct calculation using Equation (\ref{eqepsilonout}), we get that $n_{NMSD}=log_2(n)$. Therefore,
\begin{equation}
    \label{eqpszmsd2}
    p_{szMSD} \geq \dfrac{1}{n}.
\end{equation}

 One might ask, why do other MSD protocols like those of \cite{BK05,BH12}, for example, not work (using our techniques)? The answer to this question has to do with the number of noisy input $T$  states $n_{noisy}$ with fidelity $1-\varepsilon$ with respect to an ideal $T$-state, needed to distill a single $T$-state of sufficiently high fidelity $1-\varepsilon_{out}$ with respect to an ideal $T$-state. $n_{noisy}$ is usually given by \cite{BK05}
\begin{equation}
\label{eqnnoisy}
n_{noisy}=O \Big (log^{\gamma}(\dfrac{1}{\varepsilon_{out}}) \Big ).
\end{equation}
$\gamma$ is a constant which depends on the error correcting code from which the MSD protocol is derived \cite{HH18}. In the protocol of \cite{HHP+17} (as well as those in \cite{HH182}), $\gamma \sim 1$. Whereas for the Steane code for example \cite{RHG07}, which we used to distill $Y$-states in our 3D NN architecture, $\gamma >1$.  $\gamma \sim 1$ in the protocol of \cite{HHP+17} is what allowed  us to get a $p_{szMSD}$  of the form of Equation (\ref{eqpszmsd2}). On the other hand, the protocols of \cite{BK05,BH12,RHG07} have a $\gamma >1$, which leads to a lower bound of $p_{szMSD}$ which looks like $1/qp(n)$- by using similar arguments for calculating $n_{NMSD}$- where $qp(n)$ is $quasi$-$polynomial$ in $n$ (if one requires $\varepsilon_{out}=1/poly(n)$). Indeed, $N$ is proportional to $\alpha.n_{noisy}$, where $\alpha$ is the number of output $T$-states with error $\varepsilon_{out}$. Therefore, it follows that $n_{NMSD}=O(N)=O(n_{noisy})$, and that $2^{n_{NMSD}}=2^{O(n_{noisy})}$, which is a quasi-polynomial when $\gamma > 1$. This would mean, using our proof techniques, that we would need a quasi-polynomial in $n$ (which is greater than polynomial in $n$) number of $zMSD$ copies to get a succesful instance, thereby taking us out of the scope of what is considered quantum speedup \footnote{Since quantum speedup is usually defined with respect to quantum devices using polynomial quantum resources \cite{NC2000}.}. Other protocols which we could have used and could have worked are those of \cite{Jones13,HH182} which gives $\gamma \sim 1$, or that of \cite{HH18} which gives $\gamma < 1$, albeit with a huge constant overhead of $2^{58}$ qubits \cite{HH18}. 


\subsection{Proof of Equation (\ref{eqpsuccprl})}
\label{sec exp fail}
We begin by calculating $p_{fail}=1-p_{succ}.$ Suppose we have constructed $M$ copies of $zMSD$, the probability $p_{fail}$ of not getting at least $O(n^2)$ succesful instances of $zMSD$ is given by
\begin{equation}
    \label{eqappD31}
    p_{fail}=\sum_{m=0,...,O(n^2)}{M \choose O(n^2)-m } p^{O(n^2)-m}_{szMSD}(1-p_{szMSD})^{M-O(n^2)+m}.
\end{equation}
$p_{szMSD} \leq 1-p_{szMSD}$ (Equation (\ref{eqpszmsd2})),
\begin{equation}
    \label{eqappD32}
    p_{fail} \leq \sum_{m=0,...,O(n^2)}{M \choose O(n^2)-m }  (1-p_{szMSD})^{M}
\end{equation}
Taking $M > 2O(n^2),$
\begin{equation}
   \label{eqappD33} 
    \sum_{m=0,...,O(n^2)}{M \choose O(n^2)-m } \leq O(n^2) {M \choose O(n^2)}. 
\end{equation}
Replacing Equation (\ref{eqappD33}) in Equation (\ref{eqappD32}), and using Equation (\ref{eqpszmsd2}), we get
\begin{equation}
    \label{eqappD34}
    p_{fail} \leq O(n^2) {M \choose O(n^2)}(1-\dfrac{1}{n})^{M}.
\end{equation}
Also,
$${M \choose O(n^2)}<M^{O(n^2)}.$$ Replacing this in Equation (\ref{eqappD34}) we get
\begin{equation}
    p_{fail} \leq O(n^2) M^{ O(n^2)}(1-\dfrac{1}{n})^{M}.
\end{equation}
Noting that for large enough $n$
$$(1-\dfrac{1}{n})^{n} \sim \dfrac{1}{e},$$ and taking $\dfrac{M}{n}=p(n).O(n^2)$
\begin{equation}
    p_{fail} \leq O(n^2)\big(\dfrac{M}{e^{p(n)}}\big)^{O(n^2)}.
\end{equation}
Choosing $p(n) \geq log(M)=O(log(n))$, we get that $\dfrac{M}{e^{p(n)}} \leq c$, with $c<1$ a constant.
In this case, 
$$p_{fail} \leq O(n^2)c^{O(n^2)} \leq O(n^2)\dfrac{1}{e^{O(n^2)}} \sim \dfrac{1}{e^{O(n^2)}},$$ for large enough $n$.
Thus,
$$p_{succ} \geq 1-\dfrac{1}{e^{O(n^2)}}.$$
Note that for our choice of $p(n) \geq O(log(n))$, we get that $M=O(n^3)p(n) \geq O(n^3log(n)).$ This completes the proof of Equation (\ref{eqpsuccprl}).
 
 \section{The routing circuit $\overline{C_R}$} \label{APP routing}
 
\begin{figure}[H]
\begin{center}
\graphicspath{}
\includegraphics[scale=0.5]{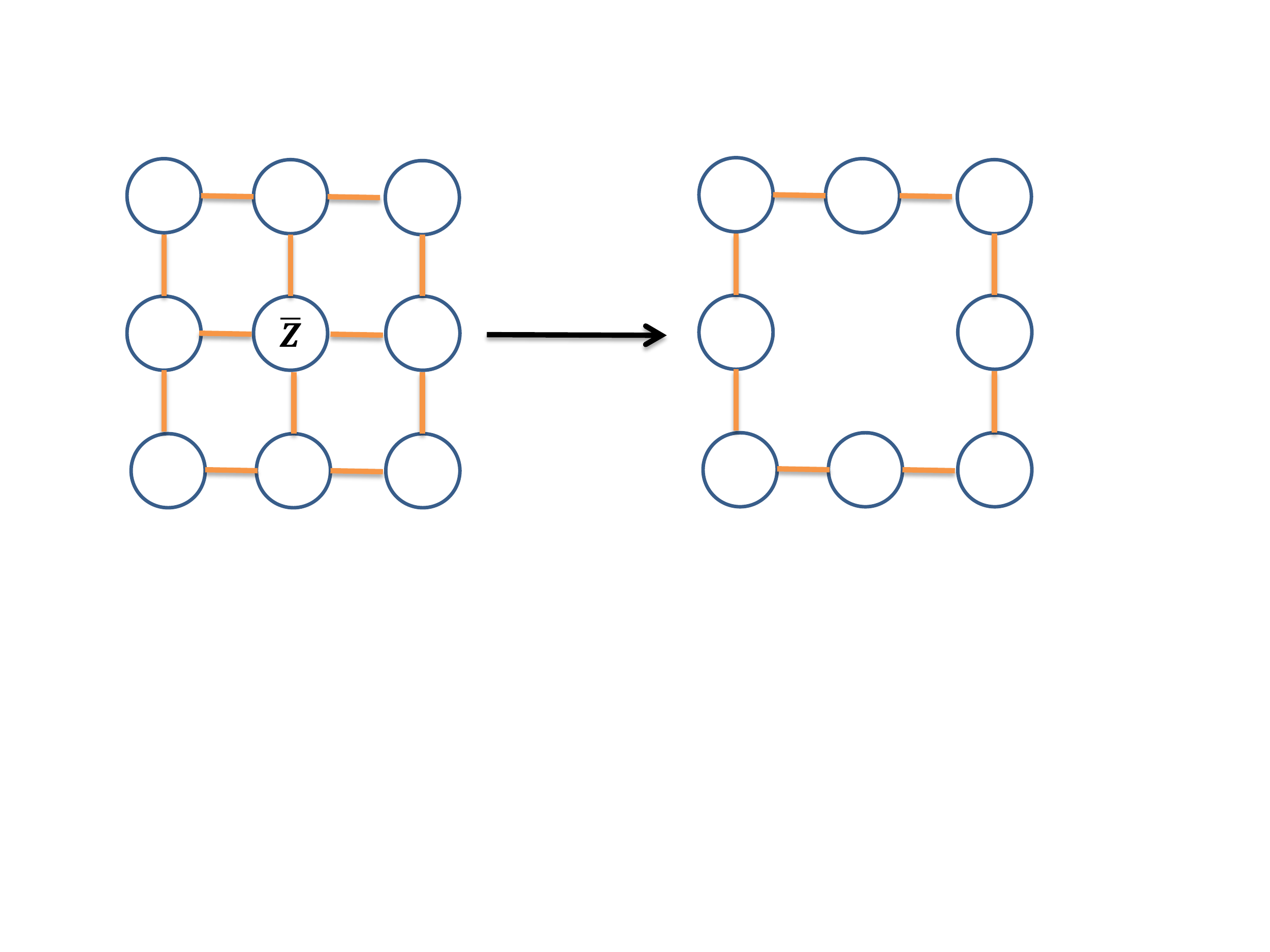}
\caption{Performing a $Z$ measurement on a vertex of a graph state removes it, up to local Pauli corrections.}
\label{FIG Z measurements on GS}
\end{center}
\end{figure}

\begin{figure}[H]
\begin{center}
\graphicspath{}
\includegraphics[ scale=0.5]{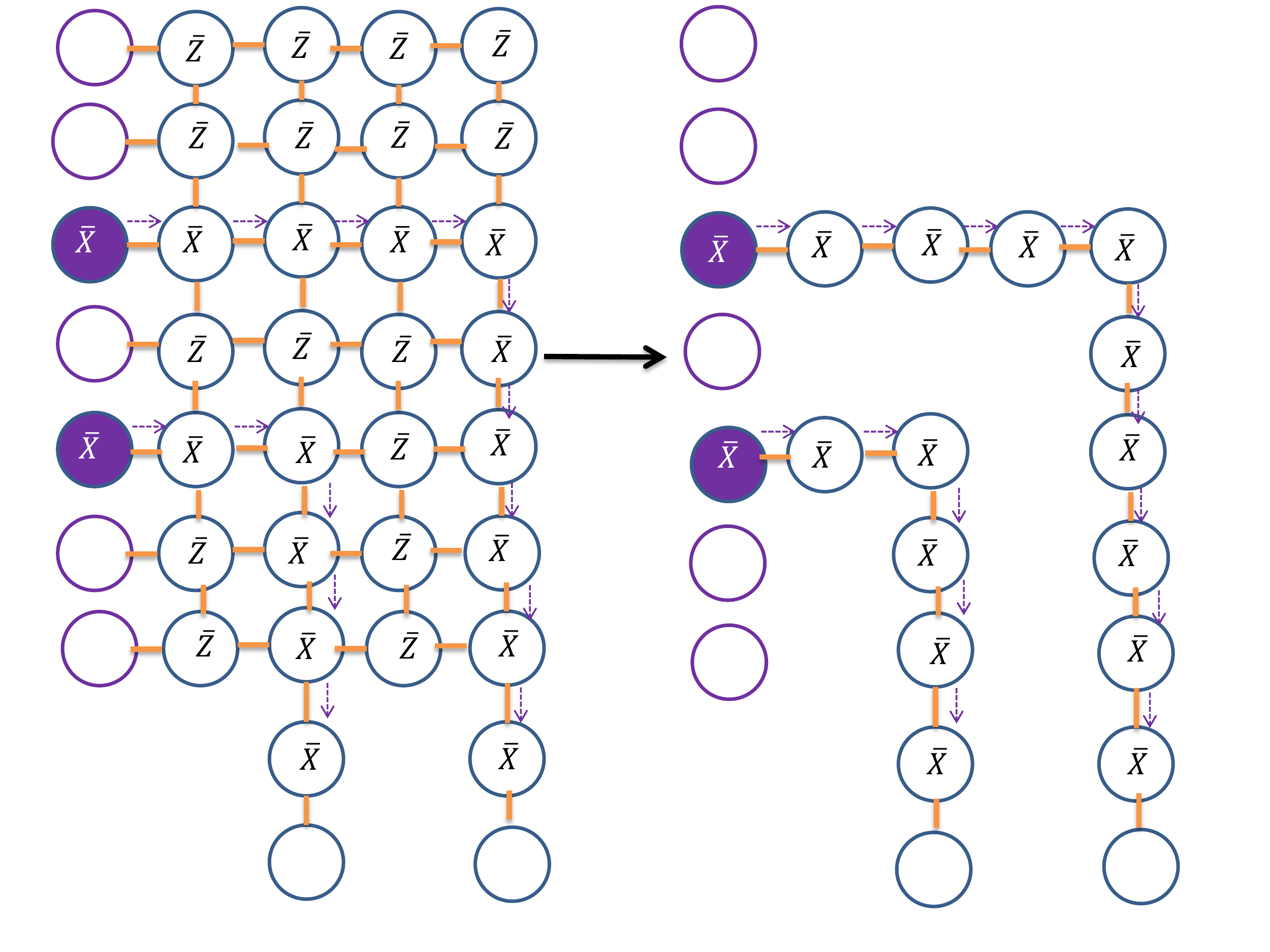}
\caption{Routing via etching out from a grid. The purple vertices on the left represent outputs of the $zMSD$. a) The filled purple vertices are identified as the successful distilled $T$ states from previous measurement results, and the paths to the outputs are identified. b) All other qubits are measured out in $Z$ and the succesful outputs are teleported via $X$ measurements.}
\label{FIG routing}
\end{center}
\end{figure}

 The main idea of the MBQC based routing is to use the fact that in a graph state, measurements allow us to etch out desired paths. In particular performing a $\overline{Z}$ measurement removes a vertex and its edges \cite{HDE+06}, as illustrated in Figure \ref{FIG Z measurements on GS}. Once a path is etched out, $\overline{X}$ measurements teleport the state along it. 
Given $m$ systems to route out of a possible $p$, a grid of size $2pm$ is sufficient for unique paths to be etched out.
An example of how this works for a grid is illustrated in Figure \ref{FIG routing} for $m=2$ and $p=7$ \footnote{Again, we can measure all qubits only in $\overline{Z}$ if we add a constant-depth layer of $\overline{H}$ gates to the graph state.}.
In our case, we have a total of $O(n^3log(n))$ outputs of all the $zMSD$, of which $O(n^2)$ will be successful, hence the number of ancilla we require scales as $O(n^5log(n))$.

It is worth explaining why the overall noise on the routed  $\overline{\rho_T}_{out}$  will still be local stochastic with constant rate. 
Firstly, note that $\overline{C_R}$ is a constant depth Clifford circuit composed of single and two-qubit Clifford gates acting on outputs of $zMSD$ circuits, and therefore all local stochastic noise after each depth one step of this circuit can be treated as a single local stochastic noise $E_{d} \sim \mathcal{N}(m)$   with constant rate $m$ at the end of this circuit, as in Equation (\ref{eq7PRL}) \cite{BGK+19}. 
The outputs of $zMSD$ circuits are acted upon by local stochastic noise with constant rate (as seen earlier overall noise on $zMSD$ is local stochastic with constant rate, therefore noise acting on a subset of qubits of $zMSD$ (the outputs) is also local stochastic with the same rate \cite{BGK+19}), and therefore can be incorporated as preparation noise (analogous to $E_{prep}$ in Equation (\ref{eq7PRL})) with $E_{d}$ to give a net local stochastic noise $E \sim \mathcal{N}(c)$ with constant rate $c$ acting on qubits of $\overline{C_R}$. 
After measurements, the unmeasured outputs of $\overline{C_R}$ will also be acted upon by $E^{'} \sim \mathcal{N}(c)$ which is local stochastic with same rate as $E$, but with smaller support, from the properties of local stochastic noise \cite{BGK+19}. 
\section{Error correction in our 3D NN architecture}
\label{appRHG1}
In this section we will show how the probability of failure $p_{fail}$ of decoding in our 3D NN architecture can be made $polynomially$  low. $p_{fail}$ here is equivalent to $1-p(ne)=1/poly(n)$  in appendix \ref{app zMSD}. Thus; obtaining $p_{fail}=1/poly(n)$ allows us to recover the same results for error correction as the 4D NN architecture. We will assume that classical postprocessing is instantaneous. We will work with local stochastic noise and, as discussed in the main text, deal with a single local stochastic noise $E \sim \mathcal{N}(q)$ which is pushed forward until after the measurements \cite{BGK+19} (see Equation (\ref{eq7PRL})). The (constant) rate $q$ satisfies $q \leq 0.0075$ \cite{RHG07}, that is, it is below the threshold of fault-tolerant computing in the RHG construction. As argued in \cite{DKL+02}, the probability $p_{fail}$ can be calculated by calculating the number of ways in which the minimal weight matching results in a non-trivial error, that is, an error  stretching across at least $L_{m}$ qubits, where $L_{m}$ is the minimum between the perimeter of the defect and  the (minimal) distance between two defects \cite{RHG07,DKL+02}. $p_{fail}$ can be calculated by using the following relation \cite{DKL+02}

\begin{equation}
    \label{eqSAP1}
    p_{fail} \leq P(n) \sum_{L \geq L_{m}} n(L).prob(L).
\end{equation}
This relation simply counts the number of ways in which a relevant non-trivial error can occur, this type of error is restricted to errors induced by self-avoiding walks (SAWs) on the lattice, as argued in \cite{DKL+02}. $n(L)=6.5^{L-1}$
calculates all possible  SAWs of total lenght $L$ originating from a fixed point in the lattice \cite{DKL+02}, $P(n)=poly(n)$ is the the total number of fixed points (i.e physical qubits) on the lattice, since SAWs can originate at any fixed point, and $prob(L) \leq (4q)^{\frac{L}{2}}$ is the probability that the minimal matching induces an error chain (SAW) of length $L$, this probability is calculated using the techniques in \cite{DKL+02}, but adapted to local stochastic noise (whereas independent depolarizing noise acting on each qubit was considered in \cite{DKL+02}). The sum is over all non-trivial errors of lenght $L_m \leq L \leq poly(n)$. Noting that 
\begin{equation}
\label{eqSAP2}
    P(n)\sum_{L \geq L_{m}} n(L).prob(L) \leq poly(n)(poly(n)-L_m)\dfrac{6}{5}(100q)^{\frac{L_{m}}{2}} \leq poly^{'}(n).(0.75)^{\dfrac{L_m}{2}} \sim \dfrac{poly^{'}(n)}{e^{0.06L_m}},
\end{equation}
where $poly^{'}(n)$ is some polynomial in $n$. Choosing $L_m=\alpha.log(n)$ with $\alpha$ a positive constant, and replacing Equation (\ref{eqSAP2}) in (\ref{eqSAP1}) we get
\begin{equation}
    \label{eqSAP3}
    p_{fail} \leq \dfrac{poly^{'}(n)}{n^{0.06\alpha}}.
\end{equation}
Finally, choosing $$0.06\alpha > deg(poly^{'}(n)),$$ and replacing this in Equation (\ref{eqSAP3})
we obtain our desired polynomially low bound for $p_{fail}$
\begin{equation}
   \label{eqSAP4}
   p_{fail} \leq \dfrac{1}{poly(n)}.
    \end{equation}

Now, we want to find an estimate of the individual rates of preparation, gate application and measurement in our 3D NN architecture. Assuming at each layer of the circuit, qubits are acted upon by a local stochastic noise $E \sim N(p)$ with $0<p<1$ a constant, we get that $q \leq 4p^{4^{-D-1}}$ \cite{BGK+19}, where $D$ is the total quantum depth of the RHG construction.  $D=6$, one step for preparation, one for (non-adaptive) measurements (assuming instantaneous classical computing as mentioned earlier), and four steps for preparing the RHG lattice \cite{RHG05}. Setting $q \leq 0.0075$ \cite{RHG07}, we get that the errors in preparation, gate application, and measurement should satisfy $p \leq \sim e^{-40000}$. Note that, for completeness, the threshold error rate for the distillation $\varepsilon$ should also be taken into account. Usually, $\varepsilon$ should be lower than some constant \cite{reichardt2005quantum} in order for distilation to be possible, but this is accounted for in the chosen value of $q$ \cite{RHG07}. 

\section{ Distillation and overhead in our 3D NN architecture}
\label{appRHG2}
\subsection{Distillation}
In this subsection, we will discuss distillation of logical $Y$-states in our 3D NN construction. The distillation of $T$-states in this construction is exactly the same as in appendix \ref{app zMSD}, but instead of using the protocol of Theorem 4.1 in \cite{HHP+17}, we use the protocol with $\gamma \sim 1$ (see appendix \ref{app zMSD}) in \cite{HH182} which allows transversal implementation of logical $T$-gates and is thus compatible with the RHG construction \cite{RHG07,RHG05}.

The distillation of $Y$-states is done with the $[7, 1, 3]$ Steane code \cite{RHG07}. This code has a $\gamma \sim log(7)/log(3) \sim 1.77$ \cite{HH18}. Therefore, the total number of logical ancilla qubits (including qubits prepared in initial noisy logical $Y$-states $\overline{\rho_{Y}}_{noisy}$) needed to distill a logical $Y$-state of fidelity $1-\varepsilon^{'}_{out}$ is given by \cite{BH12} (see appendix \ref{app zMSD})
\begin{equation}
    \label{eqdistillY1}
    N_Y=O(log^{1.77}(\dfrac{1}{\varepsilon^{'}_{out}})).
\end{equation}
Choosing $\varepsilon^{'}_{out}=1/O(poly(log(n)))$  as in the main text, we get that
\begin{equation}
    \label{eqdistillY2}
    N_Y=O(log^{1.77}(poly(log(n))) \sim O(log^{1.77}(log(n))).
\end{equation}
It is straightforward to see that, for high enough $n$,
\begin{equation}
    \label{eqdistillY3}
    N_Y < log(n).
\end{equation}
$N_Y$ can be though of as the number of logical qubits of a 2D  logical cluster state needed to distill a logical $Y$-state of fidelity $1-\varepsilon^{'}_{out}$. As in appendix \ref{app zMSD}, if we do this MBQC non-adaptively, we only succeed with probability 
\begin{equation}
    \label{eqdistillY4}
    P_s \geq \dfrac{1}{2^{N_Y}} \geq \dfrac{1}{n}.
\end{equation}
In our case, we need $O(n^5log^2(n))$ logical $Y$-states of fidelity $1-\varepsilon^{'}_{out}$  in order to distill $O(k.n)=O(n^4)$ $T$-states to be used in the construction of $\overline{C_2}$.   $O(n^5log^2(n))$ is the number of qubits of $\overline{C_1}$ when $k=O(n^3)$ (number of columns of $|G\rangle$ ). Therefore, by results in appendix \ref{sec exp fail}, we would need $\overline{C^{'}_1}$ to be composed of $O(n^6log^3(n))$ logical qubits in order to distill, with exponentially high probability of success, enough ($O(n^5log^2(n))$) logical $Y$-states with fidelity $1-\varepsilon^{'}_{out}$.

Now, we will see why logical $Y$-states of fidelity $1-\varepsilon^{'}_{out}=1-1/O(poly(log(n)))$  suffice to disill 
$O(n^5log^2(n))$ $T$-states with fidelity $1-\varepsilon_{out}=1-1/O(poly(n))$. In the construction of $\overline{C_1}$
in appendix \ref{APP subsec zMSD}, replacing a perfect logical $Y$-state with a logical $Y$-state of fidelity $1-\varepsilon^{'}_{out}$, then measuring this state, results in applying the gate $\overline{H}\overline{Z(\pi/2)}$
with probability $1/2(1-\varepsilon^{'}_{out})$ instead of $1/2$ in the perfect logical $Y$-state case. Therefore, the success probability of $zMSD$ becomes in this case
\begin{equation}
    \label{eqdistillY5}
    p_{zMSD} \geq \dfrac{1}{n}(1-\varepsilon^{'}_{out})^{O(log(n))},
\end{equation}
as compared with Equation (\ref{eqpszmsd2}) in the perfect logical $Y$ case. By choosing, as we did, $\varepsilon^{'}_{out}=1/poly(log(n))$, the above equation can be rewritten, for large enough $n$, as
\begin{equation}
    \label{eqdistillY6}
    p_{zMSD} \geq \dfrac{1}{n}(1-\dfrac{1}{O(poly(log(n)))})^{O(log(n))} \sim \dfrac{1}{n}(1-\dfrac{1}{O(poly^{'}(log(n)))}) \sim \dfrac{1}{n}.
\end{equation}
Thus, we have recovered Equation (\ref{eqpszmsd2}), and therefore can now use the same analysis as in appendix \ref{app zMSD} to distill logical $T$-states of fidelity $1-\varepsilon_{out}$ in our 3D NN construction. This will allow us to construct the sampling problem Equation (\ref{eq14PRL}) showing a quantum speedup. 

\subsection{Overhead}
In this subsection, we will estimate the overhead (number of physical qubits in the 3D RHG lattice) of our 3D NN construction. As in \cite{RHG07}, we will make use of the concept of a logical elementary cell. Each logical elementary cell is a 3D cluster state composed of $\lambda \times \lambda \times \lambda$ elementary cells (each of which has eighteen qubits). Logical elementary cells can be either primal or dual. Each logical elementary cell contains a  single defect. A defect inside a logical elementary cell has a cross section of $d \times d$ (perimeter $4d$) on any plane perpendicular to the direction of simulated time. For our purposes, we will choose $\lambda=O(d)$, and $d=O(log(n))$. This will ensure that the perimeter of the defect ($4d$) and the distance between two defects ($\lambda-d$) satisfy the conditions in appendix \ref{appRHG1}. In this picture, every logical qubit (composed of two defects of the same type) needs $2\times 18 \times \lambda^{3}=O(log^3(n))$ physical qubits. In order to not talk about primal or dual logical qubits (recall that computation is always carried out on logical qubits of same type, but we need braiding between two defects of different type in order to implement some gates such as $\overline{CNOT}$), we will assume each logical qubit needs four cells (two primal, two dual) to be defined, and therefore the number of physical qubits per logical qubit is $4 \times 18 \times \lambda^{3}=O(log^3(n))$. Now, all we need to do is calculate the number of logical qubits we need in total. Preparations of logical qubits in states $|\overline{+}\rangle$, $\overline{\rho_T}_{noisy}$, and $\overline{\rho_Y}_{noisy}$, and applying $\overline{CNOT}$ gates can be done using a constant number of intermediate elementary logical cells \cite{RHG07}. Therefore, we will only need to count the total number of logical qubit inputs for circuits $\overline{C^{'}_1}$, $\overline{C^{'}_R}$, $\overline{C_1}$, $\overline{C_R}$, and $\overline{C_2}$, then multiply this by a constant in order to get the total number of needed logical qubits including preparations and logical CNOT applications.  As already calculated in the previous subsection, the total number of logical qubits of $\overline{C^{'}_1}$ is $O(n^6log^3(n))$. The total overhead of circuits $\overline{C_1}$, $\overline{C_R}$, and $\overline{C_2}$ is $O(n^9poly(log(n))$ logical qubits, this is obtained by the same calculations as done in our 4D NN architecture, but with replacing $k=O(n)$ with $k=O(n^3)$, in order for the partially invertible universal set condition to be satisfied \cite{MGDM19}. Finally, the routing circuit $\overline{C^{'}_R}$ (see appendix \ref{APP routing}) needs $O(n^6log^3(n).n^5log^2(n))=O(n^{11}log^5(n))$, this term dominates the scaling. Multiplying $O(n^{11}log^5(n))$  by a constant (to account for preparation and logical CNOT gates overhead), then by $O(log^3(n))$ (to get the number of physical qubits), we get that the overall number of physical qubits needed is $O(n^{11}poly(log(n))$.

\end{document}